\documentclass[prd,10pt,twocolumn,showpacs,nofootinbib, aps,superscriptaddress]{revtex4-1}

\usepackage{bm}
\usepackage{graphicx}
\usepackage{multirow}
\usepackage{amsmath}
\usepackage{mathrsfs}
\usepackage{amssymb}
\usepackage{hyperref}
\usepackage{color}
\usepackage{xspace}
\usepackage{verbatim}
\usepackage{mathtools}
\usepackage{booktabs}
\usepackage[dvipsnames]{xcolor}
\usepackage[caption=false]{subfig}
\hypersetup{urlcolor=Mulberry,
	    citecolor=OliveGreen,
	    linkcolor=Blue, 
	    colorlinks=true}
\definecolor{green}{rgb}{0,0.5,0}
\definecolor{red}{rgb}{0.5,0,0}
\definecolor{blue}{rgb}{0,0,0.5}

\newcommand{\kpc}{\,\textrm{kpc}}
\newcommand{\pc}{\, \textrm{pc}}
\newcommand{\dbd}[2]{\ifmmode \frac{\textrm{d}#1}{\textrm{d}#2}\else $\textrm{d}#1/\textrm{d}#2$\fi}
\newcommand{\pbp}[2]{\ifmmode \frac{\partial#1}{\partial#2}\else $\partial#1/\partial#2$\fi}

\newcommand{\dfx}{\textrm{d}^4 x}
\newcommand{\Mpl}{M_{\rm Pl}}
\newcommand{\Msol}{\, \textrm{M}_\odot}
\newcommand{\vbf}{\mathbf{v}}
\newcommand{\xbf}{\mathbf{x}}
\newcommand{\drm}{\mathrm{d}}

\begin{document}

\title{Stellar kinematics from the symmetron fifth force in the Milky Way disk}

\author{Ciaran A. J. O'Hare}\email{ciaran.aj.ohare@gmail.com} \affiliation{Departamento de F\'isica Te\'orica, Universidad de Zaragoza, Pedro Cerbuna 12, E-50009,
Zaragoza, Espa\~{n}a}
\affiliation{School of Physics and Astronomy, University of Nottingham, University Park, Nottingham, NG7 2RD, United Kingdom}

\author{Clare Burrage}
\affiliation{School of Physics and Astronomy, University of Nottingham, University Park, Nottingham, NG7 2RD, United Kingdom}

\date{\today}
\smallskip
\begin{abstract}
It has been shown that the presence of non-minimally coupled scalar fields giving rise to a fifth force can noticeably alter dynamics on galactic scales. Such a fifth force must be screened in the Solar System but if unscreened it can have a similar observational effects as a component of non-baryonic matter. We consider this possibility in the context of the vertical motions of local stars in the Milky Way disk by reframing a methodology used to measure the local density of dark matter. By attempting to measure the properties of the symmetron field required to support vertical velocities we can test it as a theory of modified gravity and understand the behaviour of screened scalar fields in galaxies. In particular this relatively simple setup allows the symmetron field profile to be solved for model parameters where the equation of motion becomes highly nonlinear and difficult to solve in other contexts. We update the existing Solar System constraints for this scenario and find a region of parameter space not already excluded that can explain the vertical motions of local stars out to heights of 1 kpc. At larger heights the force due to the symmetron field profile exhibits a characteristic turn over which would allow the model to be distinguished from a dark matter halo.
\end{abstract}

\maketitle

\section{Introduction}\label{sec:intro}
The era of precision cosmology has confirmed that the $\Lambda$CDM paradigm is in exceptional accordance with observations from galactic scales up to the Hubble scale.  However the nature of both of the eponymous components of this model remain unknown.  We do not understand why the quantum vacuum energy contributes an effective cosmological constant that is so much larger than the observed value.  The particle nature of cold dark matter has also eluded all direct and indirect methods of detection. The masses of proposed dark matter candidates range from ultra-light scalars, $m \sim 10^{-22} \mbox{ eV}$ \cite{Hu:2000ke}, to multiple solar mass black holes $m \sim 10^{58} \mbox{ GeV}$ \cite{Bellomo:2017zsr}.

The symmetron model \cite{Hinterbichler:2010es,Hinterbichler:2011ca} (see also earlier work Refs.~\cite{Dehnen:1992rr,Gessner:1992flm,Damour:1994zq,Pietroni:2005pv,Olive:2007aj,Brax:2010gi}) is a scalar-tensor theory proposed as a way of understanding how new light degrees of freedom could be present on the largest scales in the universe, and yet remain undetected in laboratory and Solar System tests of gravity \cite{Burrage:2017qrf}. Whilst the cosmological constant problem remains unsolved it is important to understand what new physics can be present on these scales~\cite{Joyce:2014kja}.  

The symmetron scalar field is non-minimally coupled so it can be thought of as mediating a fifth force. To evade stringent Solar System constraints on such a force the model has `screening mechanism' which means it can give rise to observable deviations from general relativity (GR) on some scales, whilst remaining undetectable on others. This can be achieved through non-trivial self interactions that cause the properties of scalar perturbations (which mediate fifth forces) to depend on the background field configuration. As the scalar field is sourced by the local matter density (or equivalently by the local scalar curvature) screening ensures that the effects of the field become small in regions of high density, including the local environment, but can be larger in the cosmological vacuum.  For the symmetron model this occurs by making the strength of the coupling between the scalar field and matter dependent on the local matter energy density in such a way that the theory has spontaneous symmetry breaking between regions of high and low density. 

The effects of the symmetron mechanism on cosmological scales is the source of much interest presently, see e.g. Refs.~\cite{Llinares:2014zxa,Gronke:2015ama,Carlesi:2016yas,Gronke:2016lfd,Voivodic:2016kog,Desmond:2018euk,Ellewsen:2018tww,Mota:2018kid}. It also was recently realised that, for a particular choice of the mass of the scalar, the symmetron fifth force could play a role on sub-galactic scales \cite{Burrage:2016yjm}. In fact the presence of this force could explain the rotation curves and stability of disk galaxies without the need for any dark matter component\footnote{Galactic rotation curves were in fact explained in terms of scalar-tensor theories much earlier in Ref.~\cite{Gessner:1992flm} but in this case with the Higgs as the scalar}. Additionally the symmetron force provides a natural explanation of the observed radial acceleration relation \cite{McGaugh:2016leg} and may resolve a tension in the central velocity dispersions of globular clusters~\cite{Llinares:2018dtu} (although in this latter case a dark matter contribution is also assumed). The efficacy of screening mechanisms may also be affected if one permits scalar waves to propagate through the galaxy by finding solutions beyond the quasi-static approximation~\cite{Hagala:2016fks,Ip:2018nhl}.
 
Other attempts have been made to replace cold dark matter with a modification of gravity, including MOdified Newtonian Dynamics (MOND) \cite{Milgrom:1983pn,Bugg:2014bka}, and its extension as a Tensor-Vector-Scalar (TeVeS) theory \cite{Skordis:2009bf}, and generally covariant MOdified Gravity (MOG) \cite{Moffat:2005si,Moffat:2013sja}.  Other scalar field theories have also been constructed to reproduce galactic rotation curves \cite{Khoury:2014tka}. The interest of the symmetron is that it is a simple, well-defined, relativistic theory and yet can still reproduce observed galactic dynamics while functioning as a prototype for more complete descriptions of similar models such as in Refs.~\cite{Burrage:2016xzz,Burrage:2018dvt}. 

Of course simply explaining the rotation curves of disk galaxies is an insufficient case for considering a model an alternative to the cold dark matter paradigm. Nevertheless the symmetron remains of interest for two reasons:  Firstly, considering the symmetron as a modification of gravity, it shows that such modifications can have significant, or even dominant, effects on galactic scales and yet remain compatible with local tests. If such modifications of gravity are present this would lead to significant corrections to predictions of the amount of dark matter in our galaxy. Hence we can gain insight into how these effects manifest across the available model parameter space by considering the extreme case where it completely substitutes the role of dark matter. Secondly, since in the Einstein frame the symmetron takes the form of a light-Higgs portal theory with spontaneous symmetry breaking~\cite{Burrage:2018dvt}, we are able show that the inclusion of non-minimal couplings between a light scalar and the standard model can give rise to fifth forces playing a significant role on galactic scales.  Light scalars are currently considered as potential cold dark matter particles, and yet their non-minimal couplings to gravity are commonly neglected.  Such couplings can be tuned to be absent at one energy scale, but will be generated by quantum corrections as the theory is allowed to run \cite{Herranen:2014cua}.

In this work we consider our own galaxy, and whether the observed dynamics of local stars in the Milky Way (MW) can be explained by the presence of a fifth force. This has been attempted previously in the context of MOND~\cite{Bienayme:2009wb,Margalit:2015zla}, but here, as in Ref.~\cite{Burrage:2016yjm} we consider the symmetron model. In this particular scenario rather than finding a galactic solution in the radial direction we focus on the vertical direction. Using the kinematics of stars perpendicular to the plane of the galactic disk ($z$) one can attempt to model the shape of the modified gravitational potential provided by the symmetron field. As this is a first exploratory analysis we reduce the problem to a 1-dimensional one, with idealised but, importantly, self-consistent data.  We do this because our desire is to understand in particular the behaviour of the symmetron in the Galactic disk and its impact on baryonic matter, whilst also exploring the available model parameter space in detail. As such we require that our methods are highly efficient and that we are able to isolate the impact of the symmetron. As well as lending insight into the behaviour of scalar fields on sub-galactic scales, this simplification allows us to find solutions in highly stiff and nonlinear regimes, and demonstrate whether a follow-up analysis with, for example, the recent full release of data from the Gaia survey would be worthwhile. Using a 1-dimensional approximation does unfortunately set certain limitations on the scales over which one may probe, for example one may use only the closest stars ($z\lesssim 1\, \kpc$) so that the various approximations required are valid, for ignoring the radial motions of stars and the gradient of the MW rotation curve.

To begin in Sec.~\ref{sec:symmetron} we introduce the symmetron model and its screening mechanism and describe existing constraints on the model parameters. Then in Sec.~\ref{sec:localstars} we describe the vertical motions of local stars and how they can be used to measure the shape of the gravitational potential in the disk. We then apply the symmetron model to this astrophysical setting and in Sec.~\ref{sec:analysis} we describe the numerical routine used to solve for the symmetron field. We present our results in Sec.~\ref{sec:results} and summarise in Sec.~\ref{sec:summary}.

\section{The symmetron}\label{sec:symmetron}
Modifications of Einstein's general relativity (GR) often involve the addition of a scalar field. Such scalar-tensor theories can be written in two forms:  in the Jordan frame where the scalar couples non-minimally to gravity, and does not directly couple to matter fields; and the Einstein frame where the scalar couples non-minimally to matter and minimally to gravity.  The difference between these two frames is just a series of field redefinitions so physical observables are unaffected by the change of frame. 

 The general Einstein frame action for a conformal scalar-tensor theory with a scalar field $\varphi$, coupled to matter via some function $A(\varphi)$ is
\begin{eqnarray}
 S =& \int \dfx \sqrt{-g} \left( \frac{\Mpl^2 R}{2} - \frac{1}{2} \partial^\mu \varphi \partial_\mu \varphi - V(\varphi) \right) \nonumber \\
&+ S_m [A^2(\varphi)g_{\mu \nu} ; \psi_m] \, ,
\end{eqnarray}
where $R$ is the Ricci scalar, $g_{\mu \nu}$ is the Einstein frame metric, $V(\varphi)$ is the potential for the field and $S_m$ is the action for matter fields $\psi_m$. The equation of motion that follows from varying this action is,
\begin{equation}\label{eq:eom1}
  \Box \varphi = -\dbd{V(\varphi)}{\varphi} - \rho \dbd{A(\varphi)}{\varphi} \, ,
\end{equation}
where the trace of the energy momentum tensor has been written in terms of the conserved local density of non-relativistic matter, $\rho$.
The form of Eq.(\ref{eq:eom1}) suggests that we might define an effective potential for the field as
\begin{equation}
 \tilde{V}(\varphi) = V(\varphi) + \rho A(\varphi) \, .
\end{equation}
Scalar-tensor theories of gravity then have an associated `fifth force' felt by test particles that depends on spatial gradients in the field\footnote{This force can be derived by considering the form of the geodesic equation for particles moving on the Jordan frame metric.}
\begin{equation}\label{eq:fforce}
  \textbf{F}_\varphi = - \nabla \ln A(\varphi) \, .
\end{equation}

The existence of fifth forces is tightly constrained by tests of GR in the Solar System (see e.g. Ref.~\cite{Bertotti:2003rm}). This means that to function as a viable modification of gravity they must be somehow screened locally. This screening can proceed via a mass dependent on the local density such as in chameleon mechanism~\cite{Khoury:2003aq,Khoury:2003rn} or through a modification to the kinetic term of the scalar field used by the Vainshtein mechanism~\cite{Vainshtein:1972sx}. The symmetron mechanism however~\cite{Hinterbichler:2011ca,Hinterbichler:2010es}, restores a spontaneously broken symmetry in regions of high density in order to screen the fifth force. To do this, the simplest symmetron model has a quartic potential,
\begin{equation}
 V(\varphi) = -\frac{1}{2}\mu^2 \varphi^2 + \frac{1}{4} \lambda \varphi^4 \, ,
\end{equation}
with quadratic coupling function,
\begin{equation}
 A(\varphi) = 1+\frac{\varphi^2}{2 M^2} \, ,
\end{equation}
such that the effective potential is
\begin{equation}
 \tilde{V}(\varphi) = \frac{1}{2}\left(\frac{\rho}{M^2} - \mu^2\right) \varphi^2 + \frac{1}{4} \lambda \varphi^4 \, .
\end{equation}
The dependence of the effective  mass term on $\rho(\mathbf{x})$ means that the theory can undergo spontaneous symmetry breaking between regions of high and low density. At high densities when the mass term is positive the potential has one minimum at $\varphi = 0$ and $\mathbb{Z}_2$ symmetry. In low enough densities the sign of the mass term flips, the symmetry is broken around the new minima at $\varphi = \pm v$ where we define $v~=~\mu /\sqrt{\lambda}$ as the field vacuum expectation value (vev) in the symmetry broken regime. 

Using Eq.~\eqref{eq:fforce} the fifth force due to the symmetron felt by a small test mass depends on the field profile as
\begin{equation}
 \textbf{F}_\varphi \approx -\frac{\varphi}{M} \nabla \frac{\varphi}{M} \, .
\end{equation}
Since the model has a nonzero vev in regions of low density, any spatial variation in the field due to the presence of matter will induce a fifth force. In regions of high enough density, since the symmetry restoration moves the vev to 0, the matter coupling vanishes and the fifth force is said to be screened. The length scale for the force is the Compton wavelength $\ell \sim 1/m$ where $m$ is the symmetron mass. In regions of high density this is $m_{\rm in}^2~=~(\rho/M^2 - \mu^2)>0$ whereas in low densities expanding the potential around the new vev $\varphi = \pm v$ gives a mass of $m_{\rm out}^2~=~2(\mu^2~-~\rho/M^2)~\approx 2\mu^2$.

The symmetron model has been designed specifically to be compatible with local tests of gravity, so it should not be surprising that constraints on the parameters of the theory, $\{M,\,\mu, \,\lambda\}$, or equivalently $\{ M, \, \mu, \, v\}$, are weak \cite{Burrage:2017qrf}.  The tightest constraints on symmetrons that we consider here - those which are light on galactic scales - come from bounds on their two post-Newtonian (PPN) parameters with non-vanishing deviations from GR in the Solar System\footnote{More massive symmetrons, with Compton wavelengths of the order $\sim 1 \mbox{ cm}$, can be constrained with laboratory searches for fifth forces using atom interferometry \cite{Burrage:2016rkv,Brax:2016wjk,Jaffe:2016fsh,Burrage:2016xzz,Burrage:2017qrf}.}. In Ref.~\cite{Burrage:2016xzz}, following Ref.~\cite{Hinterbichler:2011ca}, constraints were placed on the symmetron model using PPN parameter estimates and the existing constraints from lunar laser ranging as well as time-delay experiments performed with the Cassini spacecraft~\cite{Bertotti:2003rm,Will:2005va} (see Ref.~\cite{Sakstein:2017pqi} for further refinements that will be possible in the future). These constraints assumed a galactic density profile with dark matter so we must go back to the approach taken by Ref.~\cite{Hinterbichler:2011ca} but now using field values updated for this dark matter-free context. Firstly, Cassini and lunar laser ranging data requires that we have for the PPN parameters $\gamma$ and $\beta$ respectively, 
\begin{eqnarray}\label{eq:PPNconstraint}
|\gamma - 1| &\lesssim & 10^{-5}\\
|\beta-1| &\lesssim & 10^{-4} \, .
\end{eqnarray}
For a general scalar-tensor theory these parameters are prescribed~\cite{Will:2005va}
\begin{eqnarray}
\gamma  &=& \frac{1+\omega(\varphi)}{2+\omega(\varphi)} \, \\
\beta &=& 1+\frac{1}{(3+2\,\omega(\varphi))^2(4+2\,\omega(\varphi))} \dbd{\varphi}{(A^{-2})} \, ,
\end{eqnarray}
where $\omega(\varphi)$ is the Brans-Dicke parameter,
\begin{eqnarray}
\omega(\varphi) &=& \frac{1}{2} \left[ \frac{1}{2\Mpl^2 (\textrm{d}\ln A/\textrm{d}\varphi)^2} -3  \right] \\
&\simeq& \frac{1}{2} \left[ \frac{1}{2} \left ( \frac{M^2}{\Mpl \varphi} \right)^2 - 3  \right] \ .
\end{eqnarray}
in which we have substituted in the symmetron model coupling function $A(\varphi) = 1+\varphi^2/2M^2$. For small values of $\varphi$ (i.e. in the screened regime) one can simplify locally to get,
\begin{equation}
|\gamma - 1|_0 \simeq \left( \frac{\varphi(0) \Mpl}{M^2} \right)^2  \lesssim 10^{-5} \, ,
\end{equation}
which in most cases is the more stringent of the two constraints. Since we will solve the equation of motion in terms of a dimensionless field value $\phi \equiv \varphi/v$, only then rescaling the field to meet the required strength for the force, we can define a maximum value that the rescaled $v$ can take while satisfying the constraint Eq.~\eqref{eq:PPNconstraint}, 
\begin{equation}\label{eq:PPNconstraint_vmax}
\frac{v_{\rm max}}{\Mpl} \simeq  \frac{\sqrt{10^{-5}}}{2\phi(0)} \left(\frac{M}{\Mpl}\right)^2 \, .
\end{equation}
For a given value of $M$ the allowed range for $v$ is bounded from above by this value, but also for some regions by another condition ensuring that $v<M$, as required for the predictivity\footnote{Although on this note one should keep in mind that for the very small values of $\lambda$ that must be considered for the symmetron to act over galactic scales, the couplings are necessarily very fine tuned and radiative corrections to the theory must be carefully considered. This is especially important if one wants to consider the symmetron (and fifth forces in general) in terms of sensible quantum field theories with standard model couplings. See for example Ref.~\cite{Burrage:2018dvt} for more discussion of this issue.} of an effective field theory (EFT) description for the symmetron model~\cite{Burrage:2016xzz}. We may also set a lower limit by how weak we allow the fifth force to be. The overall strength of the force relative to Newtonian gravity can be expressed as,
\begin{equation}
\alpha \equiv \left( \frac{v}{\Mpl} \right)^2 \left( \frac{\Mpl}{M} \right)^4 \, .
\end{equation}  
Since we want the symmetron to support galactic dynamics we want it to be at least as strong as gravity when unscreened, so we set $v_{\rm min}/\Mpl = (M/\Mpl)^2$.  This constraint need not necessarily be imposed directly since it should be recovered by the inability of the solution to provide the needed force in the absence of dark matter. 

\begin{figure}[t]
\includegraphics[width=0.49\textwidth]{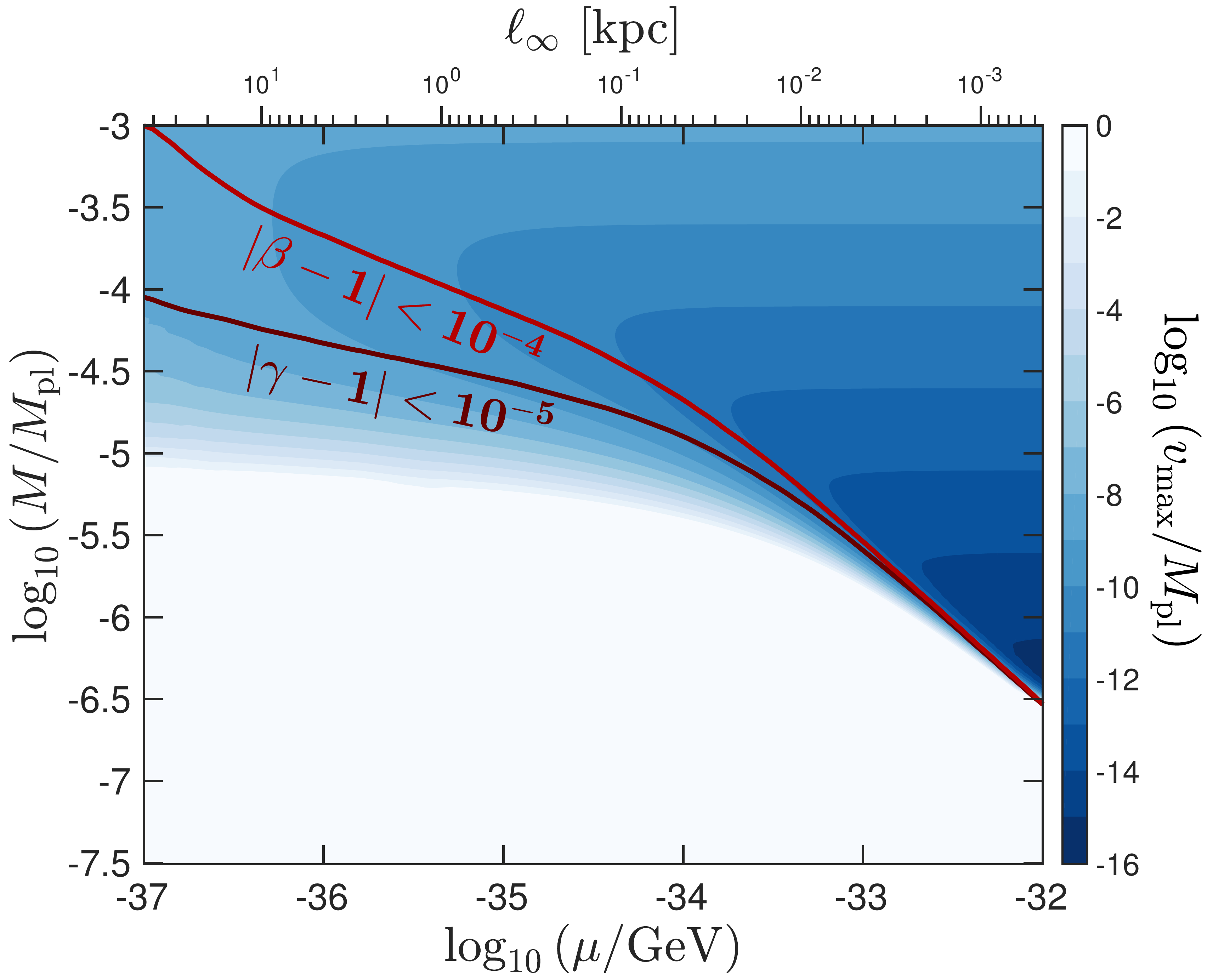}
\caption{Constraints on the symmetron parameters $\{M,\, \mu,\, v \}$ from a PPN analysis of Cassini data, expressed in analytic form in Eq.(\ref{eq:PPNconstraint_vmax}). The blue shaded regions indicate the maximum value that $v$ may take to ensure that there is no measurable fifth force in the Solar System. We also show as two red lines the boundary above which no value of $v$ that satisfies the constraints on $|\gamma-1|$ and $|\beta -1|$ can simultaneously provide a fifth force stronger than gravity, i.e. $v_{\rm max}/\Mpl<(M/\Mpl)^2$. The region below the lower line is permitted but bounded from above by $v_{\rm max}$. Note that on the top horizontal axis we show the size of the Compton wavelength in the symmetry broken regime where $\rho \approx 0$ defined as $\ell_\infty \sim 1/\sqrt{2 \mu^2}$.} \label{fig:PPN}
\end{figure}

We display the PPN constraints from measurements of $|\gamma -1|$ and $|\beta-1|$ on the three symmetron parameters in Fig.~\ref{fig:PPN}. In the $M-\mu$ plane we show there is a region that gives a value for $v_{\rm max}$ incompatible with the other bounds. Note that computing these constraints requires the knowledge of $\varphi(0)$ from the solution of the symmetron equation of motion in the Galaxy which is the subject of later sections. In the interest of clarity we introduce them here first. Figure~\ref{fig:PPN} shows the value of $v_{\rm max}$ from Eq.~\eqref{eq:PPNconstraint_vmax} that sets the upper limit for the rescaled fifth force for a given pair of $M$ and $\mu$. The regions delimited by the red lines show the boundaries above which $(v_{\rm max}/\Mpl)<(M/\Mpl)^2$, or in other words there are no values above each line for which the symmetron can provide a fifth force stronger than gravity whilst also satisfying the Solar System constraints $|\gamma-1|_0 \lesssim 10^{-5}$ and/or $|\beta-1|_0 \lesssim 10^{-4}$. Below the lower line there are values of $v$ that can satisfy all the constraints, but are bounded by the value $v_{\rm max}$ indicated by the shade of blue. In the white region the value of $v_{\rm max}$ exceeds $M$ so at this point the upper limit is set instead by the requirement for the predictivity of the EFT: $v<M$ (in Planck units). As described earlier, the bound set by Cassini for $|\gamma -1|$ is the most stringent for the range we explore here. However the line for $|\beta-1|$ becomes stronger when reaching down to smaller values of $M$ since $|\beta-1|$ contains an enhancement of $(M_{\rm Pl}/M)^2$ relative to $|\gamma -1|$.

\section{The motions of local stars}\label{sec:localstars}
The growing velocity dispersion profile for stars at increasing heights ($z$) above the plane of the MW Galactic disk demonstrates the need for a source of additional gravitational potential beyond what can be supplied by the observed distribution of baryonic matter. If we are to replace dark matter with a screened fifth force as suggested in Ref.~\cite{Burrage:2016yjm} we must show that as well as for circular motions it can mimic the role played by the dark matter halo in explaining the observed dynamics of these local stars as well . If it can then the symmetron fits at least one more requirement to be a viable replacement for dark matter on sub-galactic scales, in which case we can compare the ranges of required parameter values for $M$, $\mu$ and $v$ with those found through other means, e.g. in the Solar System, galactic rotation curves, dwarf galaxies or the CMB. 

When describing local stellar kinematics one generally must start with the collisionless Boltzmann equation,
\begin{equation}
 \dbd{f}{t} = \pbp{f}{t} + \nabla_x f \cdot \vbf - \nabla_v f \cdot \nabla_x \Phi = 0
\end{equation}
where $f(\xbf,\vbf,t)$ is the phase space distribution function for the stellar field. In the Newtonian weak-field gravity we also have Poisson's equation which relates the gravitational potential $\Phi$ to the total mass density $\rho= \rho_{\rm stars} + \rho_{\rm gas} + \rho_{\rm dm} + ...$
\begin{equation}\label{eq:poiss1}
 \nabla_x^2 \Phi = 4 \pi G \rho \,  .
\end{equation}
In cylindrical co-ordinates, $(R,\,\theta,\,z)$, one can extract three Jeans equations by integrating the Boltzmann equation over all velocities. This requires that we express the phase space distribution in terms of $\nu(\xbf)$ which is just the positional part of $f(\xbf,\vbf,t)$. Here we are interested in only one of these; the $z$-direction (referred to as the vertical direction) which is,
\begin{equation}\label{eq:zjeans}
 \frac{1}{R}\pbp{(R \nu \sigma_{R z})}{R} + \pbp{(\nu \sigma_z^2)}{z} + \nu \pbp{\Phi}{z} = 0 \, ,
\end{equation}
where the relevant components of the velocity dispersion tensor are,
\begin{equation}
 \sigma_{Rz}(\xbf) = \frac{1}{\nu(\mathbf{x})} \int \drm^3 \vbf \, f(\xbf,\vbf) \, (v_R - \langle v \rangle_R)(v_z - \langle v \rangle_z) \, ,
\end{equation}
and
\begin{equation}
 \sigma^2_z = \frac{1}{\nu(\mathbf{x})} \int \drm^3 \vbf \, f(\xbf,\vbf)\, (v_z - \langle v \rangle_z)^2 \, .
\end{equation}

The first term in Eq.(\ref{eq:zjeans}) is known as the tilt term and involves the coupling of radial and vertical motions of the stars. For a first approximation, as long as we work close enough to the disk plane we can suitably ignore this term. The reason is because for a first order expansion around $R = 8 \kpc$ and $z=0$ the gravitational potential is separable in $R$ and $z$ and hence the cross term must vanish. In reality however the tilt term it is simply just smaller than the other terms. In estimates of the local dark matter density, ignoring the tilt term adds around a 10\% uncertainty~\cite{Read:2014qva}, which is most important when the stellar density distribution is sampled beyond $\sim$ 1 kpc~\cite{Silverwood:2015hxa}. Close enough to the center of the disk it is satisfactory to use a 1-dimensional approximation and indeed this has been used frequently in the past in testbed analyses such as this. Moreover this allows us to assume the separability of the symmetron equation of motion as we discuss in Sec.~\ref{sec:analysis}. Under the approximation, the above equation is simplified to
\begin{equation}
 \pbp{(\nu \sigma_z^2)}{z} + \nu \pbp{\Phi}{z} = 0 \, ,
\end{equation}
which has a solution for the stellar density distribution,
\begin{equation}
 \frac{\nu}{\nu(0)} = \frac{\sigma_z^2(0)}{\sigma_z^2} \exp{\left(-\int_0^z \frac{1}{\sigma_z^2(z')} \pbp{\Phi}{z} \drm z'\right)} \, .
\end{equation}

We must also deal with Poisson's equation. Expanding first in cylindrical coordinates,
\begin{equation}\label{eq:poiss2}
 \pbp{{}^2\Phi}{z^2} + \frac{1}{R}\pbp{}{R}\left(R\pbp{\Phi}{R}\right) = 4 \pi G \rho \, ,
\end{equation}
where we ignore the azimuthal term since for local tracers the density is essentially symmetric in this direction. The second term on the left hand side can be written in terms of the rotation curve,
\begin{equation}
 v^2_c(R,z) = R \pbp{\Phi}{R} \, .
\end{equation}
Again, for stars sufficiently close to the plane of the disk, and at $R\simeq 8 \kpc$, the rotation curve is roughly flat which means we can safely ignore the second term of Eq.(\ref{eq:poiss2}) leaving,
\begin{equation}\label{eq:1dpoiss}
 \pbp{{}^2 \Phi}{z^2} = 4\pi G \rho \, .
\end{equation}

The baryonic part of $\rho$ is comprised of stars ($\star$), gas ($g$) and stellar remnants ($\bullet$) with a total surface density of roughly $\Sigma_b =  \Sigma_\star +\Sigma_g +\Sigma_\bullet 
\simeq (28.9+18+7.2) \Msol \,\pc^{-2}$, with most of the uncertainty on the gas~\cite{Flynn:2006tm}. Since we explore symmetron model parameters over many orders of magnitude, varying the precise shape of the baryonic density profile will not be a particularly enlightening exercise for these first tests. So we use a simple two parameter function that has been used in the past as an approximate analytic model~\cite{Read:2014qva, Silverwood:2015hxa}
\begin{equation}
 \rho_b(z) = \frac{1}{4\pi G} \bigg| \frac{k_{\rm b} d^2_{\rm b}}{(d^2_{\rm b}+z^2)^{3/2}}\bigg| \, ,
\end{equation}
where $d_{\rm b} = 0.18$~kpc and $k_{\rm b} = 1500$ (km/s)$^2$ kpc$^{-1}$. As a proxy for the inevitable uncertainty on the baryonic surface density we allow errors of $\Delta d_{\rm b} = 0.02$~kpc and $\Delta k_{\rm b} = 150$ (km/s)$^2$ kpc on the two parameters of this model, we discuss the impact of these uncertainties in Sec.~\ref{sec:results}. We are only considering a smooth density profile. This too is an approximation since at some level the density is composed partly of discrete objects that the symmetron will respond to individually. However even for our most massive symmetrons their unscreened Compton wavelengths are still larger than, or on the order of, the average separation between stars in the MW. Next, the vertical force profile is found by integrating the $z$-component of Poisson's equation,
\begin{equation}\label{eq:KDmodel}
 K_{\rm b} = -\pbp{\Phi}{z} = 4 \pi G \int \drm z \, \rho_b = -\frac{k_{\rm b}z}{\sqrt{z^2+d^2_{\rm b}}}\, .
\end{equation}
Using this model allows us to build a way of testing the symmetron with simulated data that is entirely self-consistent within the simplifying assumptions we have made. 

Next we have the dark matter contribution which serves as our comparison. In an NFW halo and in the absence of any dark disk, the dark matter density at our galactic radius can be treated as constant up to $z \sim 1$ kpc, with a total integrated surface density in the region of $\Sigma_{\rm dm} \sim 16 \Msol \pc^{-2}$. In our analysis, we add new physics but rather than modifying Poisson's equation with an additional contribution to $\rho(z)$, we instead modify the Jeans equation, with an additional contribution to the force felt by the tracer stars. We must instead solve
\begin{equation}
 \pbp{}{z}(\nu \sigma_z^2) + \nu \pbp{}{z}(\Phi + \ln A(\varphi)) = 0 \, ,
\end{equation}
and
\begin{equation}
 \pbp{{}^2\Phi}{z^2} = 4 \pi G \rho_b \, ,
\end{equation}
with the profile for $\varphi(z)$ supplied by the solution to the symmetron equation of motion,
\begin{equation}
 \pbp{{}^2 \varphi}{z^2} - (\frac{\rho}{M^2} - \mu^2)\varphi - \lambda \varphi^3 = 0 \, .
\end{equation}
The 1-dimensional approximation is applied to the equation of motion as well, where we have assumed separability in $z$. 

\begin{figure}[t]
\includegraphics[width=0.49\textwidth]{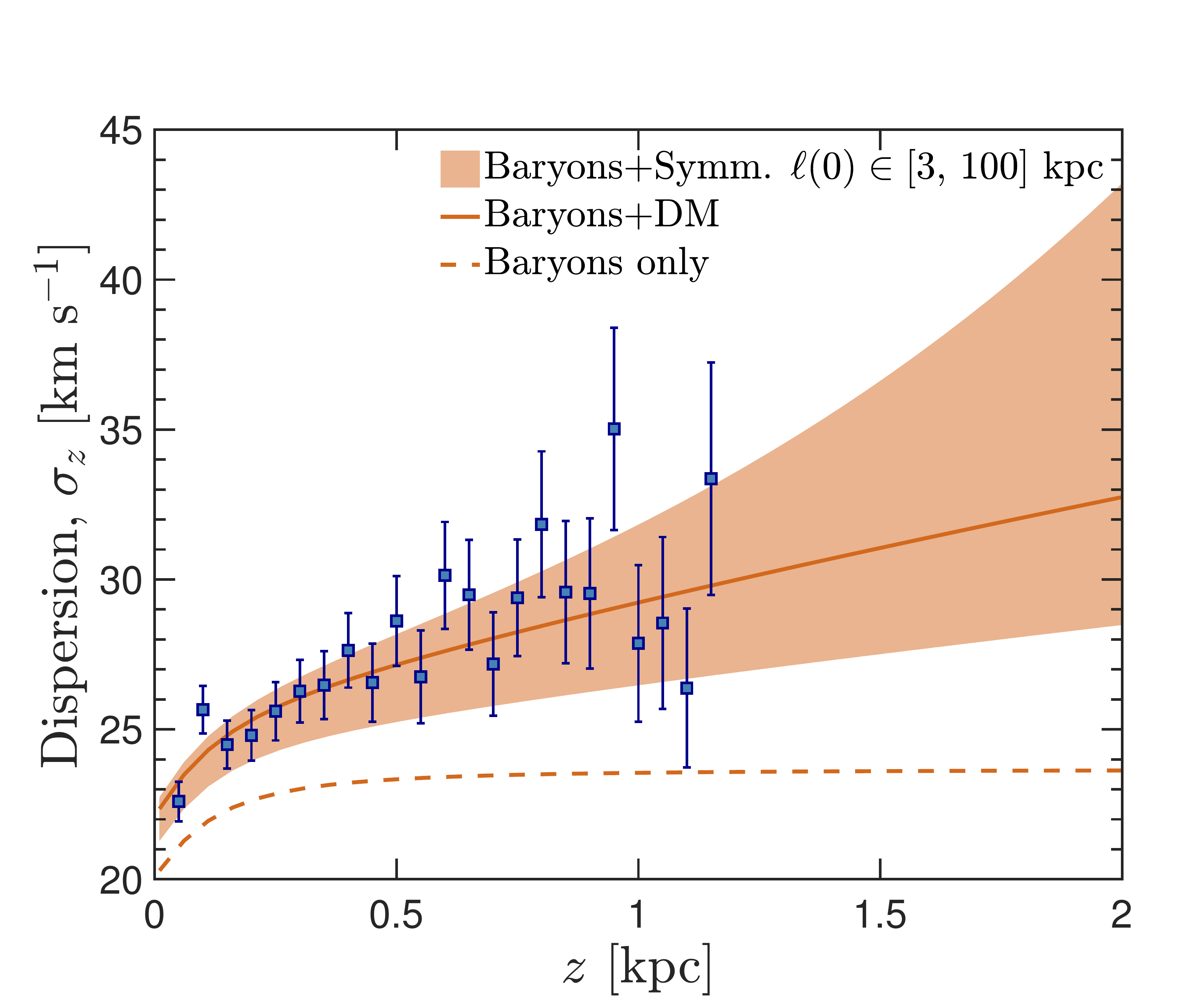}
\caption{Example data for the dispersion velocity (blue error bars) under the three parameter baryon density+dark matter model Eqs.(\ref{eq:KDmodel}) and (\ref{eq:DMmodel}). The solid line shows the underlying model (baryons+dark matter) used in the simulation, then the dashed line shows the same model with the dark matter component removed. We show as a shaded region a range of velocity dispersion profiles that can be yielded by the small-$z$ asymptotic solution to the symmetron equation of motion, for central Compton wavelengths between 3 and 100 kpc .}\label{fig:sigzdata}
\end{figure}

The data used for measurements of the local dark matter density are populations of tracer stars of a given stellar class that have enough phase space information to calculate their vertical velocity dispersion from the Jeans equation, as a function of $z$. The test of the symmetron is analogous and takes the form of a consistency check between the gravitational potential implied by the dynamics of the stars with the dynamics of the stars implied by the symmetron field profile. As a first step we can determine whether the symmetron mechanism is capable of mimicking dark matter in this way by considering the idealised scenario. The tracers we use are mock data based on this simplified model presented in Ref.~\cite{Read:2014qva}. They therefore represent something of a best case scenario, being entirely self-consistent within this model and having no observational uncertainties on the data itself which would make our conclusions less coherent. In this first instance this will be the clearest way to determine how successfully the symmetron can reproduce local stellar kinematics. A measurement would of course then be possible through the same treatment using a suitable population of stellar tracers from the high statistics/high quality Gaia data. We leave this for future work although since we explore the symmetron model space over so many orders of magnitude the precise shape of the baryon surface density profile and data uncertainties will have a subdominant impact. Hence we anticipate that many of the broad conclusions we present here will hold.

The mock data set is generated assuming a constant dark matter contribution to $\rho$, meaning the vertical force profile is linear,
\begin{equation}\label{eq:DMmodel}
K_{\rm dm}(z) = - 2 F z \,  ,
\end{equation}
where $F = 2 \pi G \rho_{\rm dm} = 267.65$ (km/s)$^2$ kpc$^{-2}$. We assume we have one isothermal tracer density profile,
\begin{equation}
\nu(z) = \nu(0) \exp\left(-\frac{z}{z_0} \right) \, ,
\end{equation}
where $z_0 = 0.4$ kpc is the scale height of the isothermal. In general the data may be comprised of a sum of many different isothermal profiles. The dispersion velocity $\sigma_z$ can be calculated once the vertical force profiles have been obtained,
\begin{eqnarray}\label{eq:sigz}
 \sigma^2_z(z) &=& \frac{1}{\nu(z)}\int_0^z \nu(z') \left(K_{\rm b}(z') + K_{\rm sym}(z')\right) \textrm{d}z'\nonumber\\ &+& \frac{\sigma^2_z(0)\nu(0)}{\nu(z)} \, .
\end{eqnarray}
We show the mock data along with the underlying model in Fig.~\ref{fig:sigzdata}. We show both the contribution from the baryons only and the baryon+dark matter model used to create the data. We require the symmetron to mimic this linearly increasing dark matter contribution. However since the data depend upon the integral of the vertical force profiles, the shape of $K_{\rm sym}$ may be significantly different from $K_{\rm dm}$ while yielding similar shapes for $\sigma_z$. We also show the small-$z$ asymptotic behaviour for a particular set of symmetron parameters. We show how these parameters and the asymptotic solution is found in the next section.

\section{Symmetron in the disk}\label{sec:analysis}
\subsection{Asymptotic solutions}
\begin{figure}[t]
\includegraphics[width=0.49\textwidth]{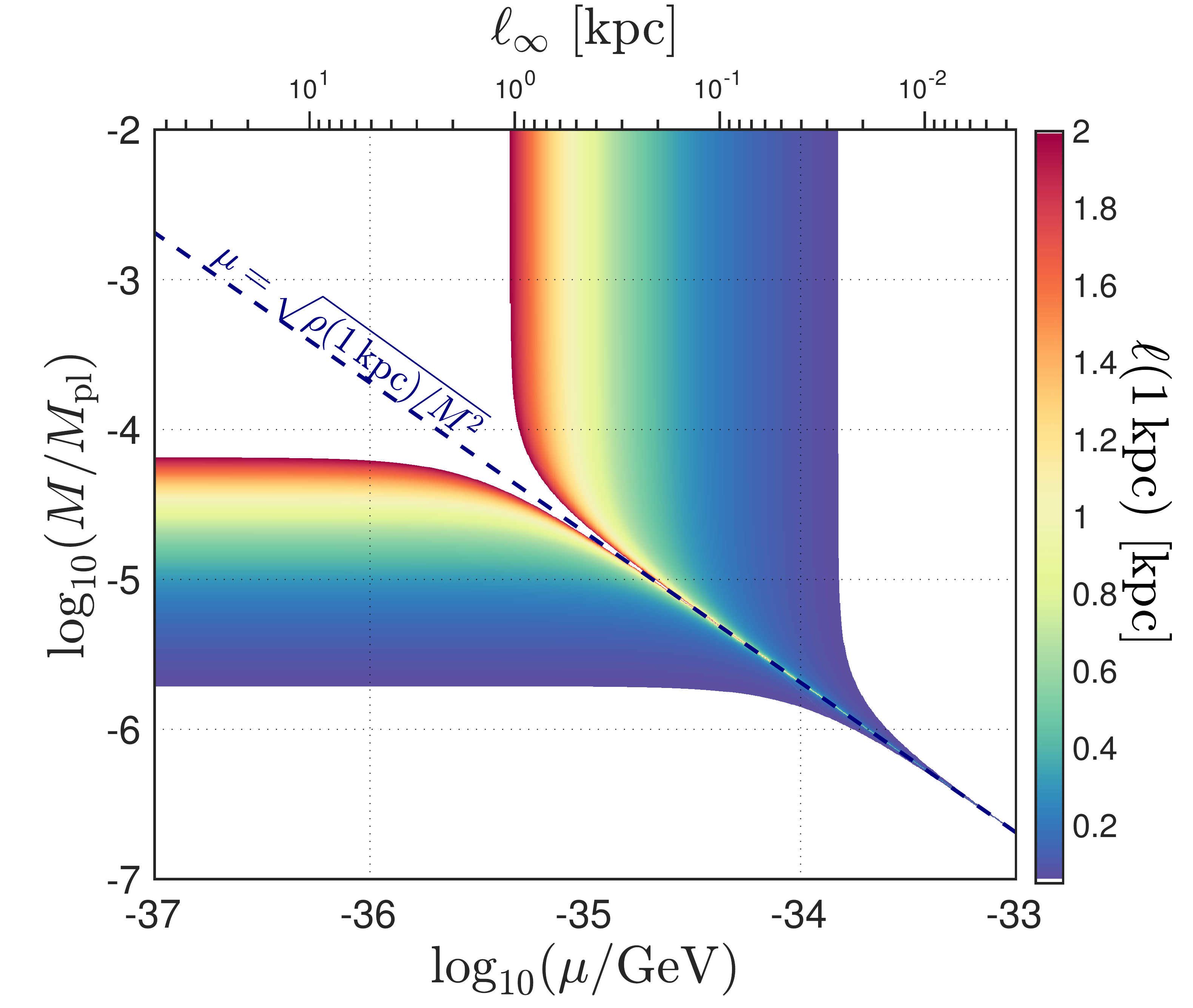}
\caption{Compton wavelength at $z = 1$ kpc as a function of model parameters $M$ and $\mu$. We limit the colour scale to lengths between 0.2 and 2 kpc where we expect the symmetron to be able to mimic a linearly increasing vertical force in the range below $z\sim 1$ kpc. The dashed line shows the values of $M$ and $\mu$ where the Compton wavelength diverges at $z=1$ kpc, since $\ell$ can grow very large inside the disk in this region we expect the small-$z$ asymptotic solution to appear here also.}\label{fig:compton1kpc}
\end{figure}

To find the fifth force experienced by the stars we need to know how the symmetron field responds to a given matter density profile $\rho(z)$. The field profile will be determined by the equation of motion,
\begin{equation}
 \pbp{{}^2 \varphi}{z^2} - \left(\frac{\rho(z)}{M^2} - \mu^2\right)\varphi - \lambda \varphi^3 = 0 \, .
\end{equation}
The full solution to this nonlinear equation must be found numerically enforcing the relevant boundary conditions. We wish to subject the equation to the conditions $\varphi'(0) = 0$ (by symmetry about $z=0$) as well as $\varphi(z)|_{z\rightarrow\infty} = v$. We can find the asymptotic behaviours of the solution at small and large $z$ by making a quadratic approximation of the potential. This is valid because at very low densities the field value will be close to $\varphi \sim v$ and at high densities close to 0 allowing us to safely ignore the $(\varphi -v)^4$ and $\varphi^4$ terms respectively in an expansion of $V(\varphi)$. For small $z$ and high densities this leads to
\begin{equation}
  \pbp{{}^2 \varphi}{z^2} - \left(\frac{\rho(z)}{M^2} - \mu^2\right)\varphi = 0 \, ,
\end{equation}
whereas for large $z$ and low densities we must solve,
\begin{equation}
  \pbp{{}^2 \varphi}{z^2} - 2\mu^2\varphi = 0 \, ,
\end{equation}
assuming that the density profile decays towards large values of $z$. Both equations yield exponential solutions. For the $z\sim 0$ case, the requirement that gradient must vanish leaves us with a solution with one free constant,
\begin{equation}\label{eq:symmprofile_smallz}
 \varphi (z) = A\, \cosh\left(\sqrt{\frac{\rho(z)}{M^2}-\mu^2}\, z_{\rm min}\right) \, .
\end{equation}
On the other hand, the field at large $z$ must decay towards the vev so we have the solution,
\begin{equation}
 \varphi(z) = v + Be^{-\sqrt{2\mu^2}z} \, .
\end{equation}
We can eliminate these constants to end up with formulae at some minimum and maximum $z$,
\begin{equation}
 \frac{\varphi'(z_{\rm min})}{\varphi(z_{\rm min})} = \sqrt{\frac{\rho(z_{\rm min})}{M^2}-\mu^2} \tanh\left(\sqrt{\frac{\rho(z_{\rm min})}{M^2}-\mu^2}\, z_{\rm min}\right) \, ,
\end{equation}
and,
\begin{equation}
 \frac{\varphi'(z_{\rm max})}{v-\varphi(z_{\rm max})} = \sqrt{2\mu^2} \, .
\end{equation}
We can proceed now by recalling that we want the fifth force to mimic dark matter, i.e. $K_{\rm sym}~\simeq~-~2Fz$. This form for the fifth force can be realised, but only for small enough $z$. Since $\varphi$ must increase with $z$ and the gradient of the field approaches 0 at both the center of the disk and at infinity, this means that the fifth force must have an extremum at some $z$. The location of this extremum will reflect the scale height of the baryonic density profile but will also be controlled by the length scale over which the symmetron responds to changes in density, i.e. the Compton wavelength. Hence the manner in which we can engineer the symmetron to best mimic dark matter will be to tune the model parameters to delay the onset of the turn over in the gradient of $\varphi$ to value of $z$ larger than the spatial extent sampled by the data. Starting from the symmetron profile at small $z$ we can use the desired scale for the force $F$ to fix the constant $A$. From the formula for the symmetron fifth force we find,
\begin{eqnarray}
 K_{\rm sym} &=& -\frac{c^2A^2}{2M^2} \sinh\left(2z\sqrt{\frac{\rho}{M^2}-\mu^2}\,\right) \nonumber \\ 
&\times& \left( \frac{z \rho'}{2 M^2 \sqrt{\frac{\rho}{M^2} - \mu^2}} + \sqrt{\frac{\rho}{M^2} - \mu^2}\right) \, .
\end{eqnarray}
So we require the symmetron coupling scale and Compton wavelength at small $z$ to roughly follow,
\begin{equation}
F \approx \frac{c^2 v^2 A^2}{2 M^2 \ell^2(z)} \, ,
\end{equation}
for $\ell(0 - 1\, {\rm kpc})>$ 1 kpc, in order for the small-$z$ solution to persist out past the extent of the data. Assuming this solution, the constant $v$ can then be freely chosen to bring the force up to the required strength. In Fig.~\ref{fig:sigzdata} we showed the resulting dispersion velocity from this field profile, assuming it extends over the full $z$ range shown. The band demonstrates the range of profiles for values of $\ell(1\,{\rm kpc}) \in [3, 100]$ kpc, and allowing a 50\% uncertainty in the initial gradient $F$. The regime of values where we expect this solution to be found is shown in Fig.~\ref{fig:compton1kpc} where we display how the Compton wavelength at 1 kpc varies with $M$ and $\mu$.

\subsection{Full solution}
\begin{figure}[t]
\includegraphics[width=0.45\textwidth]{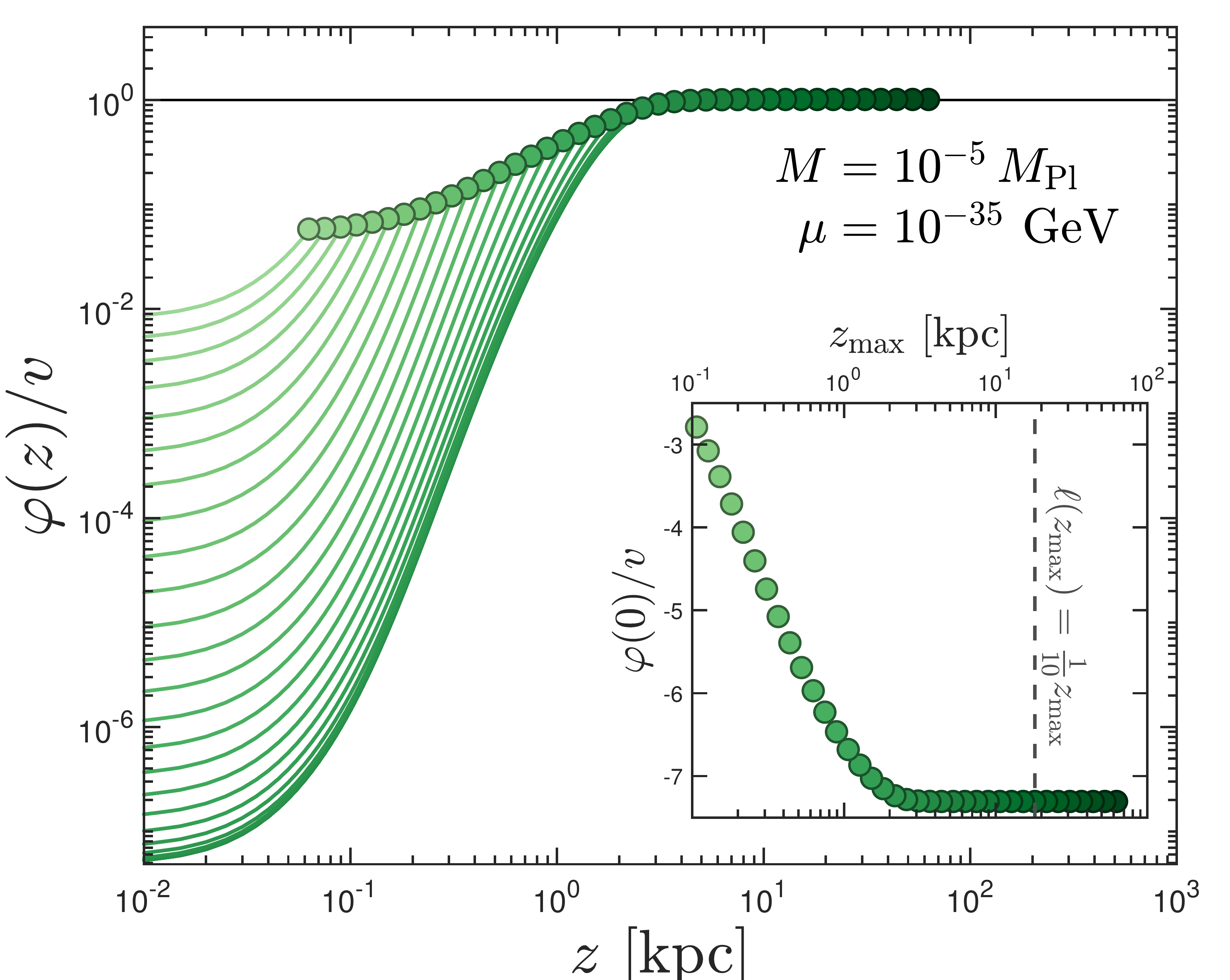}
\caption{{\bf Main:} Field profile (scaled by $v$) for $M = 10^{-5}\,\Mpl$ and $\mu = 10^{-35}$~GeV. Lines are shaded from light to dark in order of increasing values of $z_{\rm max}$ marking the upper limit of the integration domain. {\bf Inset:} The value of $\varphi/v$ at $z=0$ for this same range, we indicate the value corresponding to the solution to the equation $\ell({z_{\rm max}}) = \frac{1}{10} z_{\rm max}$ with a vertical grey dashed line. This condition approximately marks the point at which the solution is no longer sensitive to the choice of $z_{\rm max}$, as utilised in later results.}\label{fig:zmax}
\end{figure}
\begin{figure*}[t]
\includegraphics[width=0.99\textwidth]{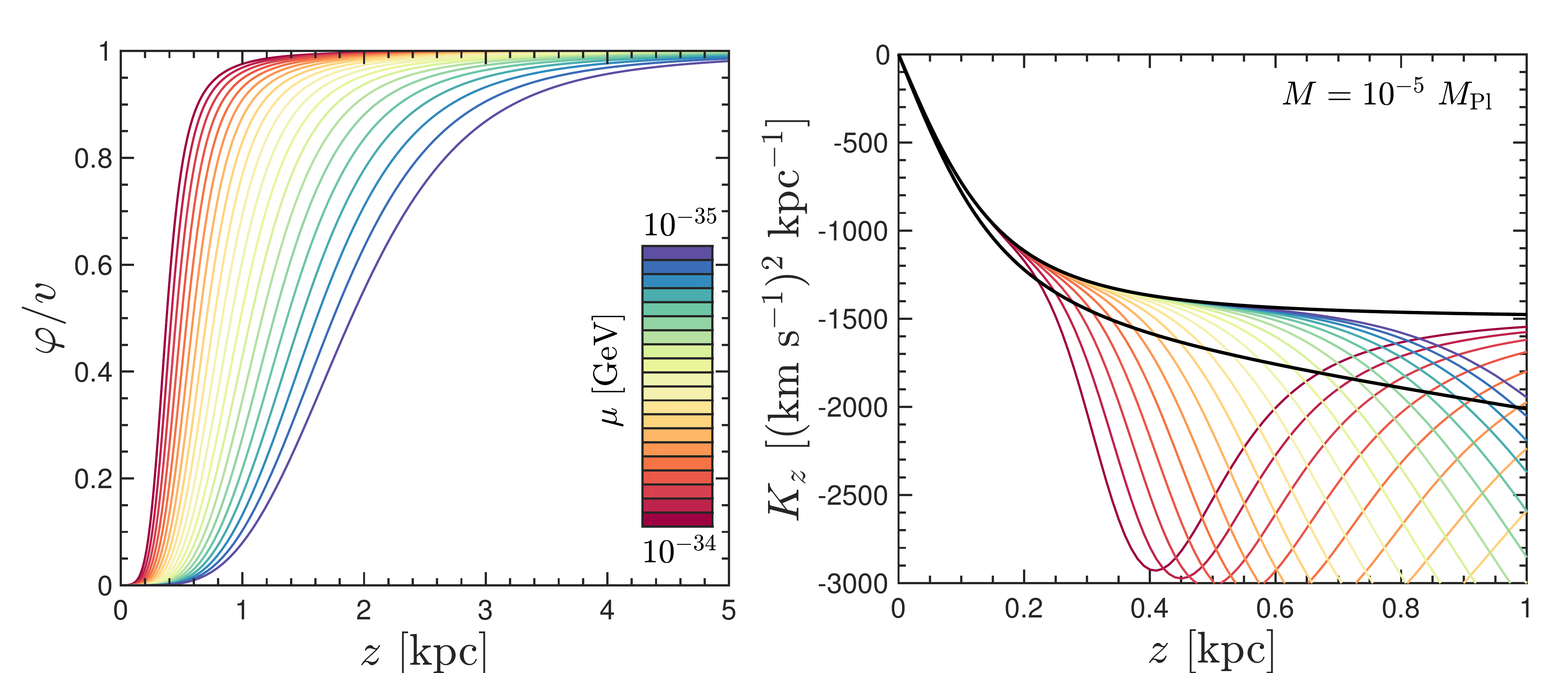}
\caption{{\bf Left:} A small subset of field profiles $\varphi(z)$ for a range of logarithmically spaced values of $\mu$ between $10^{-35}$ GeV (blue) to $10^{-34}$ GeV (red). The coupling scale is set to $M = 10^{-5}\,\Mpl$. {\bf Right:} The profile of the fifth force $K_z(z)$ up to $z=1$ kpc for the same set of inputs. The upper black line indicates the force due to baryons alone and the lower black line the force from baryons plus the dark matter model used in generating the data.}\label{fig:symmetronfield}
\end{figure*}
To solve the equation of motion numerically we work in dimensionless units where $\phi = \varphi/v$ and $x = z/{\rm 1\, kpc}$. By substituting in $\lambda =  \mu^2/v^2$ we see that $v$ can be eliminated, since this parameter only controls the scale of the field value and not the profile. The equation is then in the form,
\begin{equation}
\frac{1}{(1\, {\rm kpc})^2}\pbp{^2 \phi}{x^2}  = \frac{\rho}{M^2} \phi + \mu^2(\phi^3 - \phi)  \, .
\end{equation}
The equation of motion contains a nonlinearity and a range of length scales which can be separated by many orders of magnitude, or can conspire to have $\rho(z)/M^2 \approx \mu^2$ such that the Compton wavelength diverges within the region of interest. We also have an infinite integration domain. We find that the most efficient and reliable way to solve the equation over the large range of model parameters we need to explore is by spatially discretising the problem and dynamically relaxing the equation from some initial guess. The equation is a  Laplace-Poisson PDE, which we have already greatly simplified by moving to a 1-dimensional version of the problem. The second derivative can be computed after a suitable discretisation of $\phi_i$ with the finite difference operator,
\begin{equation}
\frac{\partial^2 \phi}{\partial x^2} = \frac{-\phi_{i+2} + 16\phi_{i+1} - 30\phi_{i} + 16\phi_{i-1} - \phi_{i-2}}{12 \Delta x^2} + \mathcal{O}(\Delta x^4) \, .
\label{eq:laplacianapprox}
\end{equation} 
Substituting this formula into the equation of motion and rearranging for $\phi_i$ we can use neighbouring points in one particular solution for $\phi$ to evolve each point in the discretisation. There are three methods one can iterate to evolve from some initial solution. The most aggressive way is to evaluate a new $\phi_i$ at each point $i$ using gradients computed from the previous iteration. However when the equation is very stiff or if the initial input solution is far from the true solution, this method is likely to diverge quickly. A more stable method is to employ a Gauss-Seidel sweep by beginning the evaluation at one end of the integration regime then computing updated gradients for each point sequentially. We find that sweeping from $z=0$ up to some $z_{\rm max}$ achieves faster convergence than the reverse. However for cases when $z_{\rm max}$ is much larger than the inner 1 kpc we can adaptively find a finer solution for small values $z$ more quickly sweeping in the reverse direction. We also implement an adaptive under-relaxation where only a fraction of the updated $\phi_i$ values are incremented if the routine begins to diverge.

Since the true integration domain is infinite in extent we must find a suitably large $z_{\rm max}$ such that the evolution at smaller $z$ is robust against changes to this value.  We can make a guess towards the size of the domain required from the behaviour of the Compton wavelength. For values of $\mu$ and $M$ for which $\ell$ grows with $z$, we find that the final solution is unaffected by changes in $z_{\rm max}$ so long as the $z_{\rm max} \gtrsim \ell(z_{\rm max})$, this is demonstrated in Fig.~\ref{fig:zmax}. Since for some choices of $M$ and $\mu$ we can have the situation in which $\ell(z) \rightarrow \infty$ for some $z$, this method for finding $z_{\rm max}$ does not always give an integration domain large enough since there will be an expanse of smaller values of $z$ where the field responds very slowly to changes in $\rho$. To rectify this we perform the checks that $z_{\rm max}$ is large enough as we explore the parameter space, ensuring that the domain always captures the full evolution. 

To begin the routine we also need an initial guess input solution. We can quite comfortably interpolate between our large and small $z$ asymptotic solutions with a guess of the form $\varphi = 1-(1-A)\,{\rm sech}^n(z/z_g)$. For much of the parameter space this function is a very good approximation to the true solution so using the PDE to perform a fit for some $\{A,n, z_g\}$ we can obtain some initial conditions that are close to the true solutions. This saves on the number of iterations needed as well as improving stability in the early stages of the relaxation. Unfortunately however for some of the very stiff areas of the parameter space this guess profile is insufficient because there can be a large separation between the scale over which the field initially increases from its central value and the scale needed for the field to decay to $v$. This intermediate, highly nonlinear regime is difficult to capture with the sech profile. However for the type of analysis we are developing here, we must solve the equation of motion many times for a range of model parameters. So when we scan across the parameter space, we can start from a point where the equation is unstiff and easy to solve (at large $M$ and $\mu$) with the aforementioned guess, then move gradually towards the stiff part of the parameter space (small $M$ and small $\mu$) using each previous solution as the input for the next. This too greatly saves in terms of the number of iterations required to find each individual solution and also reduces the risk that the routine will diverge.

Another problematic region is for very small values $M$ when the $\rho(z)/M^2$ term dominates over a large range of $z$ forcing $\varphi$ to be remain very close to 0, up to some height much larger than usual where the field begins to decay towards $v$. Generally this occurs for coupling scales below $M\lesssim10^{-6} \Mpl$. The numerical issue is due to the fact that the method for finding $z_{\rm max}$ from the behaviour of the Compton wavelength does not account for this nonlinear thin-shell effect. But by extending the integration domain by the length over which this effect persists we can reach accurate solutions as $M$ decreases. Also since the field approaches values extremely close to 0, we can resolve precision issues in calculating gradients by fitting to the analytic $\varphi \propto \cosh(z/\ell)$ solution which becomes an extremely good approximation in this regime.

\section{Results}\label{sec:results}
\begin{figure*}[t]
\includegraphics[width=0.49\textwidth,trim={0.4cm 0 1cm 0},clip]{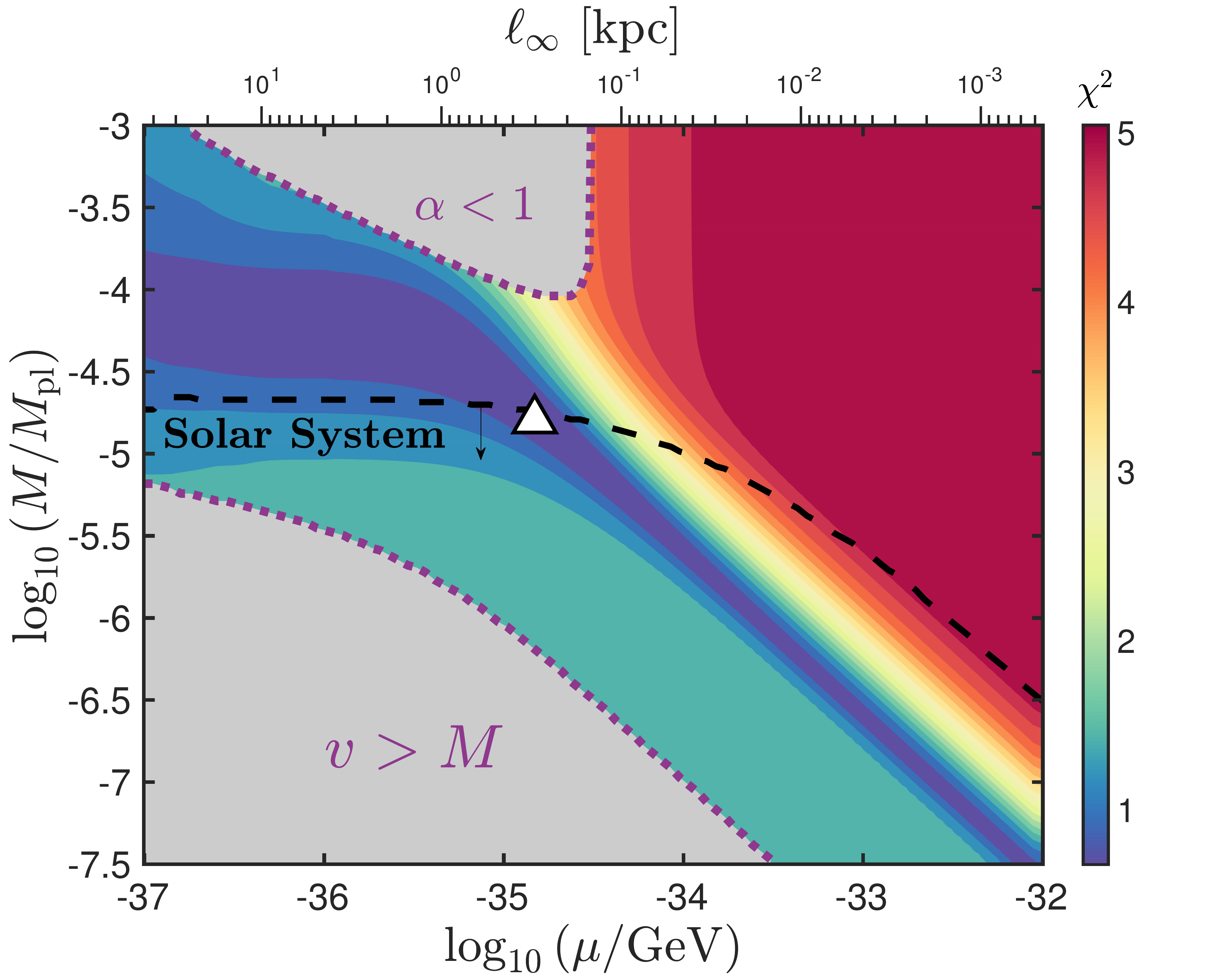}
\includegraphics[width=0.44\textwidth,trim={0 0 5cm 0},clip]{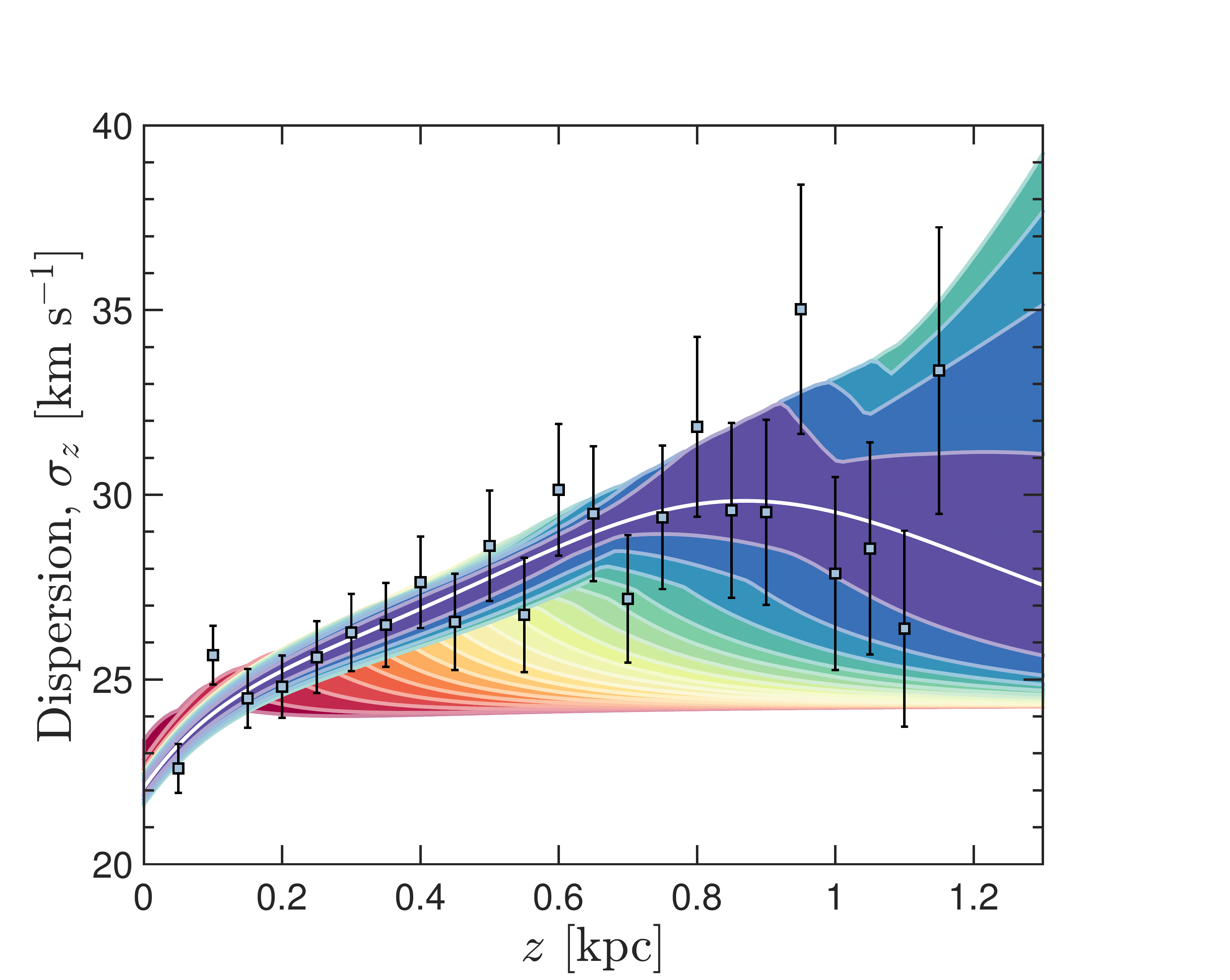}
\caption{{\bf Left:} Reduced $\chi^2$ value for the fit to $\sigma_z$ data, as a function of the symmetron parameters $M$, $\mu$ and $v$. The grey regions mask values of $\mu$ and $M$ for which the best fit value of $v$ results in a fifth force that weaker than gravity: $\alpha<1$, or is incompatible with the predictivity of the EFT description: $v>M$. We also show as a dashed black line the boundary above which $v$ will always lead to a measurable fifth force in the Solar System, with the region below this line being permitted by the Cassini measurement. {\bf Right:} Fit velocity dispersion profiles for the same range of values of $M$, $\mu$ and $v$. The colours correspond to $\chi^2$ values according to the same scale. The white line is the dispersion velocity profile for the point labelled by a white triangle in the left hand panel.}\label{fig:sigzfit}
\end{figure*}

We show the partial evolution of the symmetron field in the left hand panel of Fig.~\ref{fig:symmetronfield} for a range of values of $\mu$ and, simply for visual clarity, a single value of $M$. Then in the right hand panel of Fig.~\ref{fig:symmetronfield} we show the profile of the fifth force Eq.\eqref{eq:fforce}, as a function of $z$ for the same parameter ranges as in the previous figure. We also display as black lines the vertical force due to the baryonic density alone, and the fifth force from baryons and a constant dark matter contribution, which was used in generating the data that serves as our comparison.

The result of the fit to dispersion velocity data is shown in Fig.~\ref{fig:sigzfit}. The fit is performed by first solving the symmetron equation of motion for a given set of $\{M,\,\mu\}$. We then use the symmetron field profile to calculate the vertical force according to equation Eq.~\eqref{eq:fforce}. Then inserting this into Eq.~\eqref{eq:sigz} we calculate the expected dispersion velocities, however we need to set the scale of the strength of the force by choosing $v$. We do this by implementing a simple $\chi^2$ test comparing the calculated $\sigma_z(z)$ with the data from Fig.~\ref{fig:sigzdata}. We minimise the $\chi^2$ value as a function of $v$. In the left hand panel we display the reduced $\chi^2$ value as a function of $M$ and $\mu$, with regions with a fitted value of $v$ incompatible with the constraints overlain. We emphasise not to place too much weight on the numerical value of $\chi^2$ here since the fit takes the values of $k_{\rm b}$ and $d_{\rm b}$ as perfectly known, we simply wish to visually display how the shapes of the symmetron induced velocity dispersion profiles compare across the model space and which are incompatible with the bounds. In addition we show the resulting fits for $\sigma_z(z)$ in the right hand panel, not including the solutions for parameter values ruled out (indicated by the grey regions). There is a band of symmetron parameters can reproduce the dispersion velocity data up to around $z\sim 1$~kpc, importantly one can notice that this band has values of $\varphi$ at the solar position that allow the model to escape the PPN bounds for the fitted value of $v$. It was suggested in Ref.~\cite{Burrage:2016xzz} that this degenerate band of solutions might be present and indeed one could exploit this to find solutions for symmetrons with even smaller Compton wavelengths\footnote{However the symmetron response to individual stars may become important for larger values of $\mu$ than displayed here.}. Although there will be a limitation on this from the quickly decreasing constraint on $|\beta -1|$ for very small values of $M$. The model that provides the best fit we indicate with a white triangle in the left hand panel, and a white line on the right hand side. These values are $M = 10^{-4.8}\,\Mpl$, $\mu = 10^{-34.9}$ GeV and $v = 10^{-8.3}\,\Mpl$, which only just escapes the bound, however all values within the darkest blue region provide a similar shape at these heights. Interestingly the symmetron can provide a decent fit at level of $\sigma_z(z)$ even though the dark matter and symmetron vertical forces have quite different profiles, as shown in Fig.~\ref{fig:symmetronfield}. We anticipate that including the shape of the baryon density in $r$ and using stars at larger heights above the disk including the tilt term may allow the two models to be distinguished, if this characteristic turn over in the vertical force can be measured (or otherwise). 

Although we do not discuss this idea in detail, one might add a further constraint on the parameter space here by enforcing the stability of the galactic disk. This was shown in Ref.~\cite{Burrage:2016yjm}, and amounts to a bound of (very approximately) $v\gtrsim 6\times 10^{-3} \, M$ which our best fit region partly satisfies. However the precise numerical factor is uncertain and depends upon on an integral of the field over the full Galaxy which is not provided in our results. Nevertheless since our field profiles vary over a significantly larger range of values of $\varphi$ than the assumption used in the calculating this constraint, the numerical factor would be much larger bringing our results much more into consistency with disk stability. 

\begin{figure}[t]
\includegraphics[width=0.45\textwidth]{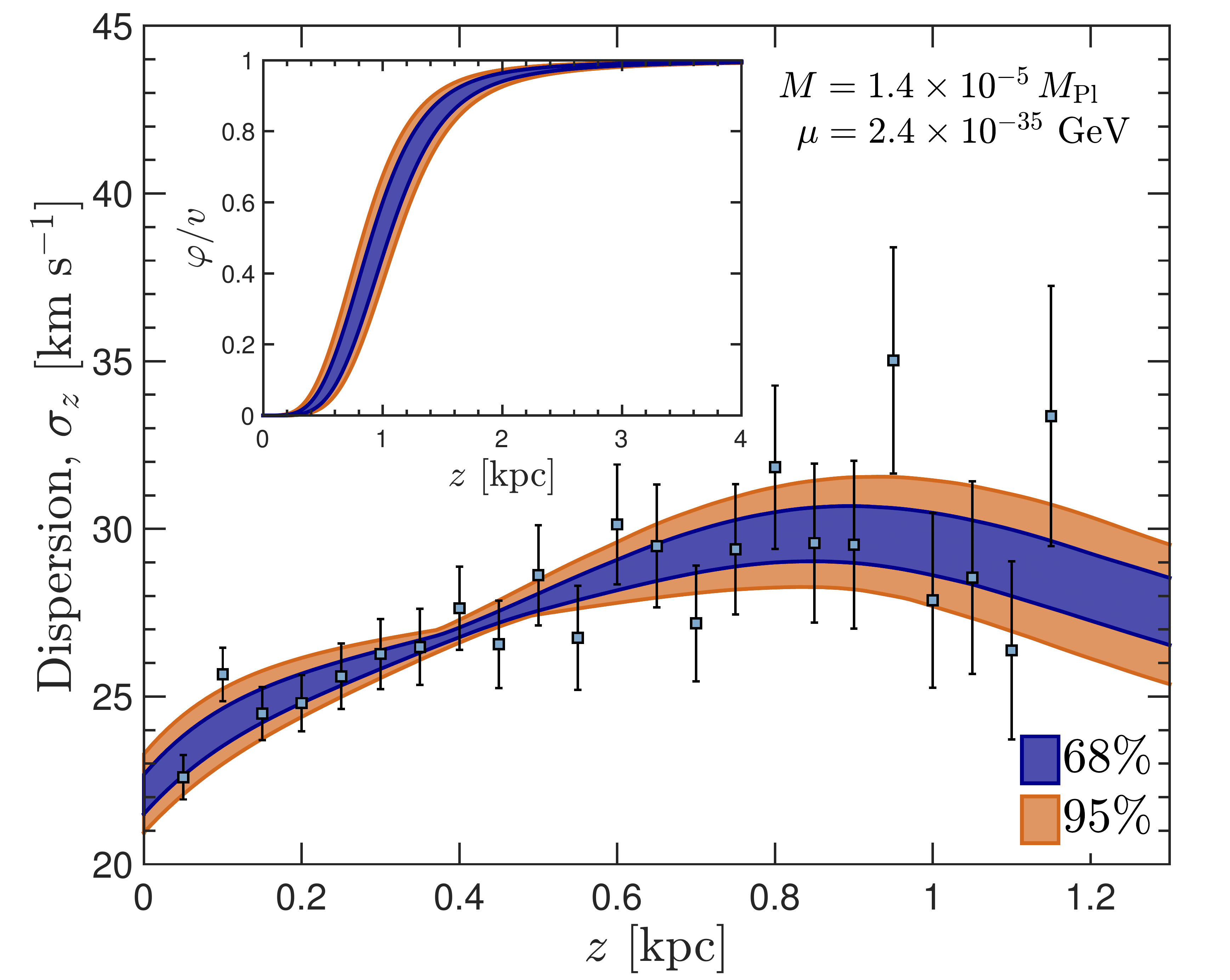}
\caption{{\bf Main:} Range of dispersion velocity profiles when varying the two baryon density parameters $k_{\rm b}$ and $d_{\rm b}$ within 68\% and 95\% uncertainties as described in the text. {\bf Inset:}  Corresponding field profiles $\varphi/v$.}\label{fig:sigzfitKD}
\end{figure}

The above results hold for a specific input baryon density model, finally for comparison we demonstrate the effect of uncertainties in this profile by varying the two parameters $k_{\rm b}$ and $d_{\rm b}$ within approximate 68\% uncertainties of $\Delta k_{\rm b}\sim 150$ (km s$^{-1}$)$^2$ kpc$^{-1}$ and $\Delta d_{\rm b} \sim 0.02$ kpc respectively. We show how this affects the field profile and resulting best fit dispersion profile in Fig.~\ref{fig:sigzfitKD}, when values of $k_{\rm b}$ and $d_{\rm b}$ are set as unknown parameters and fit with the above uncertainties.  Allowing the baryon density to vary we are able to recover a similar fit but with the addition of an uncertainty in the recovered $\sigma_z(z)$ equivalent to a change of factor of a few in the parameters $M$ or $\mu$ describing the shape of the profile, or up to a factor of two in $v$ setting the strength of the force. We find that if one has a sharper decrease in $\rho(z)$ due to a smaller value of $d_{\rm b}$ or a higher peak density due to a larger value of $k_{\rm b}$ then these both have the effect of increasing the strength of the force and hence lowering the required value of $v$. Thus we expect that one may permit a widening of our best fit region (as well as a weakening of the Solar System constraint) by up to a factor of two if one treats the baryonic density profile as not perfectly well known. Although we have implemented this uncertainty in a rather simplistic way (a full likelihood fit including these parameters would be a more appropriate method when using a real data set) it nevertheless serves as a heuristic indicator for the effect of changing the baryon density. As suggested earlier we see that this uncertainty plays a subdominant role when exploring over many orders of magnitude in the symmetron parameter space.

\section{Summary}\label{sec:summary}
We have found solutions for the symmetron field profile in the Milky Way disk in a simplified 1-dimensional system which allows us to explore regimes of coupling values not already explored in previous work. This also allows us to update Solar System PPN constraints for the symmetron dark matter model and find viable parameter space not already ruled out. In previous analyses this region was left unexplored due to significant numerical challenges found in solving the extremely stiff nonlinear equation of motion containing a wide range of length scales. However with a specialised dynamical relaxation method detailed in Sec.~\ref{sec:analysis} we have shown that exploring this region is beneficial since solutions here can allow the symmetron fifth force to replicate the effects of a roughly constant dark matter density with regards to the vertical kinematics of nearby stars. This is one step in constructing a model for scalar field dark matter as has already been initiated by previous groups. However we must emphasise there are a number of additional follow up questions that must be answered.

The next step will be to extend beyond the 1-dimensional system. This is crucial for two reasons. Firstly, for the symmetron parameter $\mu$ which controls the exponential decay of the field to the unscreened vev, we are forced to not explore much smaller values because one cannot trust the 1-dimensional approximation when the field decays so far away from the scale height of the disk. Already here for values of $\mu\lesssim 10^{-36}$ GeV it is likely that the 3-dimensional shape of the disk and the stellar bulge will begin to become important in computing the relevant field values at our location. However our results already identify an interesting part of the parameter space in which to look when extending to a multidimensional solver, and indeed many of our viable solutions act over short enough scales for this to not be an important issue. However the second reason we should look towards this extension is the fact that at large values of $z$ the data become sensitive to the correlations between the radial and vertical motions with the increasing importance of the tilt term in the Jeans equation. Modelling the tilt term is required for accuracy below the $\sim$10\% level in measurements of the dark matter density~\cite{Read:2014qva}, and ignoring it is responsible for a large underestimate in the values of $\sigma_z(z)$ at heights above $\sim 1$ kpc. Given that the recent Gaia-DR2 possesses very high statistics for local stellar kinematics, we anticipate strong constraints to come when folding this new data into these types of analyses. Finally we note that although the symmetron can be shown to mimic dark matter within the scope of the sample data presented here, there are notable differences in the predictions of the dispersion profile out to larger $z$ that distinguish it as a model. In particular the symmetron model exhibits a turn over in the fifth force at some $z$ as the field decays back towards the vev at low densities, and with kinematic data at large distances combined with a multidimensional solution and analysis it will certainly be possible to distinguish this prediction for galactic dynamics, further excluding more of the parameter space. Our numerical solving routine is easily extendible to more than one dimensions, it simply requires some computational power and a suitable region of parameter space in which to look, such as the region we have discovered here. Additionally we hope that analyses can take place in other astronomical contexts, for instance in other galaxies, as well as in dark matter dominated dwarfs. Evidence from the CMB and large scale structure notwithstanding, solutions in other environments will also be essential for the symmetron to evolve into a usable model of dark matter, and may indeed be the simplest method to rule it out.

\acknowledgments
CB is supported by a Leverhulme Trust Research Leadership Award and a Royal Society University Research Fellowship. CAJO has received support from a Leverhulme Trust Research Leadership Award and the grant FPA2015-65745-P from the Spanish MINECO and European FEDER. The authors are grateful for guidance from P. Millington in the early stages of the project as well as enlightening discussions and correspondence with E. J. Copeland, B. Elder, C. Llinares, and D. Saadeh. 

\bibliography{SymmetronLocalStars.bib}

\begin{thebibliography}{48}%
\makeatletter
\providecommand \@ifxundefined [1]{%
 \@ifx{#1\undefined}
}%
\providecommand \@ifnum [1]{%
 \ifnum #1\expandafter \@firstoftwo
 \else \expandafter \@secondoftwo
 \fi
}%
\providecommand \@ifx [1]{%
 \ifx #1\expandafter \@firstoftwo
 \else \expandafter \@secondoftwo
 \fi
}%
\providecommand \natexlab [1]{#1}%
\providecommand \enquote  [1]{``#1''}%
\providecommand \bibnamefont  [1]{#1}%
\providecommand \bibfnamefont [1]{#1}%
\providecommand \citenamefont [1]{#1}%
\providecommand \href@noop [0]{\@secondoftwo}%
\providecommand \href [0]{\begingroup \@sanitize@url \@href}%
\providecommand \@href[1]{\@@startlink{#1}\@@href}%
\providecommand \@@href[1]{\endgroup#1\@@endlink}%
\providecommand \@sanitize@url [0]{\catcode `\\12\catcode `\$12\catcode
  `\&12\catcode `\#12\catcode `\^12\catcode `\_12\catcode `\%12\relax}%
\providecommand \@@startlink[1]{}%
\providecommand \@@endlink[0]{}%
\providecommand \url  [0]{\begingroup\@sanitize@url \@url }%
\providecommand \@url [1]{\endgroup\@href {#1}{\urlprefix }}%
\providecommand \urlprefix  [0]{URL }%
\providecommand \Eprint [0]{\href }%
\providecommand \doibase [0]{http://dx.doi.org/}%
\providecommand \selectlanguage [0]{\@gobble}%
\providecommand \bibinfo  [0]{\@secondoftwo}%
\providecommand \bibfield  [0]{\@secondoftwo}%
\providecommand \translation [1]{[#1]}%
\providecommand \BibitemOpen [0]{}%
\providecommand \bibitemStop [0]{}%
\providecommand \bibitemNoStop [0]{.\EOS\space}%
\providecommand \EOS [0]{\spacefactor3000\relax}%
\providecommand \BibitemShut  [1]{\csname bibitem#1\endcsname}%
\let\auto@bib@innerbib\@empty
\bibitem [{\citenamefont {Hu}\ \emph {et~al.}(2000)\citenamefont {Hu},
  \citenamefont {Barkana},\ and\ \citenamefont {Gruzinov}}]{Hu:2000ke}%
  \BibitemOpen
  \bibfield  {author} {\bibinfo {author} {\bibfnamefont {W.}~\bibnamefont
  {Hu}}, \bibinfo {author} {\bibfnamefont {R.}~\bibnamefont {Barkana}}, \ and\
  \bibinfo {author} {\bibfnamefont {A.}~\bibnamefont {Gruzinov}},\ }\href
  {\doibase 10.1103/PhysRevLett.85.1158} {\bibfield  {journal} {\bibinfo
  {journal} {Phys. Rev. Lett.}\ }\textbf {\bibinfo {volume} {85}},\ \bibinfo
  {pages} {1158} (\bibinfo {year} {2000})},\ \Eprint
  {http://arxiv.org/abs/astro-ph/0003365} {arXiv:astro-ph/0003365 [astro-ph]}
  \BibitemShut {NoStop}%
\bibitem [{\citenamefont {Bellomo}\ \emph {et~al.}(2018)\citenamefont
  {Bellomo}, \citenamefont {Bernal}, \citenamefont {Raccanelli},\ and\
  \citenamefont {Verde}}]{Bellomo:2017zsr}%
  \BibitemOpen
  \bibfield  {author} {\bibinfo {author} {\bibfnamefont {N.}~\bibnamefont
  {Bellomo}}, \bibinfo {author} {\bibfnamefont {J.~L.}\ \bibnamefont {Bernal}},
  \bibinfo {author} {\bibfnamefont {A.}~\bibnamefont {Raccanelli}}, \ and\
  \bibinfo {author} {\bibfnamefont {L.}~\bibnamefont {Verde}},\ }\href
  {\doibase 10.1088/1475-7516/2018/01/004} {\bibfield  {journal} {\bibinfo
  {journal} {JCAP}\ }\textbf {\bibinfo {volume} {1801}},\ \bibinfo {pages}
  {004} (\bibinfo {year} {2018})},\ \Eprint {http://arxiv.org/abs/1709.07467}
  {arXiv:1709.07467 [astro-ph.CO]} \BibitemShut {NoStop}%
\bibitem [{\citenamefont {Hinterbichler}\ and\ \citenamefont
  {Khoury}(2010)}]{Hinterbichler:2010es}%
  \BibitemOpen
  \bibfield  {author} {\bibinfo {author} {\bibfnamefont {K.}~\bibnamefont
  {Hinterbichler}}\ and\ \bibinfo {author} {\bibfnamefont {J.}~\bibnamefont
  {Khoury}},\ }\href {\doibase 10.1103/PhysRevLett.104.231301} {\bibfield
  {journal} {\bibinfo  {journal} {Phys. Rev. Lett.}\ }\textbf {\bibinfo
  {volume} {104}},\ \bibinfo {pages} {231301} (\bibinfo {year} {2010})},\
  \Eprint {http://arxiv.org/abs/1001.4525} {arXiv:1001.4525 [hep-th]}
  \BibitemShut {NoStop}%
\bibitem [{\citenamefont {Hinterbichler}\ \emph {et~al.}(2011)\citenamefont
  {Hinterbichler}, \citenamefont {Khoury}, \citenamefont {Levy},\ and\
  \citenamefont {Matas}}]{Hinterbichler:2011ca}%
  \BibitemOpen
  \bibfield  {author} {\bibinfo {author} {\bibfnamefont {K.}~\bibnamefont
  {Hinterbichler}}, \bibinfo {author} {\bibfnamefont {J.}~\bibnamefont
  {Khoury}}, \bibinfo {author} {\bibfnamefont {A.}~\bibnamefont {Levy}}, \ and\
  \bibinfo {author} {\bibfnamefont {A.}~\bibnamefont {Matas}},\ }\href
  {\doibase 10.1103/PhysRevD.84.103521} {\bibfield  {journal} {\bibinfo
  {journal} {Phys. Rev.}\ }\textbf {\bibinfo {volume} {D84}},\ \bibinfo {pages}
  {103521} (\bibinfo {year} {2011})},\ \Eprint {http://arxiv.org/abs/1107.2112}
  {arXiv:1107.2112 [astro-ph.CO]} \BibitemShut {NoStop}%
\bibitem [{\citenamefont {Dehnen}\ \emph {et~al.}(1992)\citenamefont {Dehnen},
  \citenamefont {Frommert},\ and\ \citenamefont {Ghaboussi}}]{Dehnen:1992rr}%
  \BibitemOpen
  \bibfield  {author} {\bibinfo {author} {\bibfnamefont {H.}~\bibnamefont
  {Dehnen}}, \bibinfo {author} {\bibfnamefont {H.}~\bibnamefont {Frommert}}, \
  and\ \bibinfo {author} {\bibfnamefont {F.}~\bibnamefont {Ghaboussi}},\ }\href
  {\doibase 10.1007/BF00674344} {\bibfield  {journal} {\bibinfo  {journal}
  {Int. J. Theor. Phys.}\ }\textbf {\bibinfo {volume} {31}},\ \bibinfo {pages}
  {109} (\bibinfo {year} {1992})}\BibitemShut {NoStop}%
\bibitem [{\citenamefont {Gessner}(1992)}]{Gessner:1992flm}%
  \BibitemOpen
  \bibfield  {author} {\bibinfo {author} {\bibfnamefont {E.}~\bibnamefont
  {Gessner}},\ }\href {\doibase 10.1007/BF00645239} {\bibfield  {journal}
  {\bibinfo  {journal} {Astrophys. Space Sci.}\ }\textbf {\bibinfo {volume}
  {196}},\ \bibinfo {pages} {29} (\bibinfo {year} {1992})}\BibitemShut
  {NoStop}%
\bibitem [{\citenamefont {Damour}\ and\ \citenamefont
  {Polyakov}(1994)}]{Damour:1994zq}%
  \BibitemOpen
  \bibfield  {author} {\bibinfo {author} {\bibfnamefont {T.}~\bibnamefont
  {Damour}}\ and\ \bibinfo {author} {\bibfnamefont {A.~M.}\ \bibnamefont
  {Polyakov}},\ }\href {\doibase 10.1016/0550-3213(94)90143-0} {\bibfield
  {journal} {\bibinfo  {journal} {Nucl. Phys.}\ }\textbf {\bibinfo {volume}
  {B423}},\ \bibinfo {pages} {532} (\bibinfo {year} {1994})},\ \Eprint
  {http://arxiv.org/abs/hep-th/9401069} {arXiv:hep-th/9401069 [hep-th]}
  \BibitemShut {NoStop}%
\bibitem [{\citenamefont {Pietroni}(2005)}]{Pietroni:2005pv}%
  \BibitemOpen
  \bibfield  {author} {\bibinfo {author} {\bibfnamefont {M.}~\bibnamefont
  {Pietroni}},\ }\href {\doibase 10.1103/PhysRevD.72.043535} {\bibfield
  {journal} {\bibinfo  {journal} {Phys. Rev.}\ }\textbf {\bibinfo {volume}
  {D72}},\ \bibinfo {pages} {043535} (\bibinfo {year} {2005})},\ \Eprint
  {http://arxiv.org/abs/astro-ph/0505615} {arXiv:astro-ph/0505615 [astro-ph]}
  \BibitemShut {NoStop}%
\bibitem [{\citenamefont {Olive}\ and\ \citenamefont
  {Pospelov}(2008)}]{Olive:2007aj}%
  \BibitemOpen
  \bibfield  {author} {\bibinfo {author} {\bibfnamefont {K.~A.}\ \bibnamefont
  {Olive}}\ and\ \bibinfo {author} {\bibfnamefont {M.}~\bibnamefont
  {Pospelov}},\ }\href {\doibase 10.1103/PhysRevD.77.043524} {\bibfield
  {journal} {\bibinfo  {journal} {Phys. Rev.}\ }\textbf {\bibinfo {volume}
  {D77}},\ \bibinfo {pages} {043524} (\bibinfo {year} {2008})},\ \Eprint
  {http://arxiv.org/abs/0709.3825} {arXiv:0709.3825 [hep-ph]} \BibitemShut
  {NoStop}%
\bibitem [{\citenamefont {Brax}\ \emph {et~al.}(2010)\citenamefont {Brax},
  \citenamefont {van~de Bruck}, \citenamefont {Davis},\ and\ \citenamefont
  {Shaw}}]{Brax:2010gi}%
  \BibitemOpen
  \bibfield  {author} {\bibinfo {author} {\bibfnamefont {P.}~\bibnamefont
  {Brax}}, \bibinfo {author} {\bibfnamefont {C.}~\bibnamefont {van~de Bruck}},
  \bibinfo {author} {\bibfnamefont {A.-C.}\ \bibnamefont {Davis}}, \ and\
  \bibinfo {author} {\bibfnamefont {D.}~\bibnamefont {Shaw}},\ }\href {\doibase
  10.1103/PhysRevD.82.063519} {\bibfield  {journal} {\bibinfo  {journal} {Phys.
  Rev.}\ }\textbf {\bibinfo {volume} {D82}},\ \bibinfo {pages} {063519}
  (\bibinfo {year} {2010})},\ \Eprint {http://arxiv.org/abs/1005.3735}
  {arXiv:1005.3735 [astro-ph.CO]} \BibitemShut {NoStop}%
\bibitem [{\citenamefont {Burrage}\ and\ \citenamefont
  {Sakstein}(2017)}]{Burrage:2017qrf}%
  \BibitemOpen
  \bibfield  {author} {\bibinfo {author} {\bibfnamefont {C.}~\bibnamefont
  {Burrage}}\ and\ \bibinfo {author} {\bibfnamefont {J.}~\bibnamefont
  {Sakstein}},\ }\href@noop {} {\  (\bibinfo {year} {2017})},\ \Eprint
  {http://arxiv.org/abs/1709.09071} {arXiv:1709.09071 [astro-ph.CO]}
  \BibitemShut {NoStop}%
\bibitem [{\citenamefont {Joyce}\ \emph {et~al.}(2015)\citenamefont {Joyce},
  \citenamefont {Jain}, \citenamefont {Khoury},\ and\ \citenamefont
  {Trodden}}]{Joyce:2014kja}%
  \BibitemOpen
  \bibfield  {author} {\bibinfo {author} {\bibfnamefont {A.}~\bibnamefont
  {Joyce}}, \bibinfo {author} {\bibfnamefont {B.}~\bibnamefont {Jain}},
  \bibinfo {author} {\bibfnamefont {J.}~\bibnamefont {Khoury}}, \ and\ \bibinfo
  {author} {\bibfnamefont {M.}~\bibnamefont {Trodden}},\ }\href {\doibase
  10.1016/j.physrep.2014.12.002} {\bibfield  {journal} {\bibinfo  {journal}
  {Phys. Rept.}\ }\textbf {\bibinfo {volume} {568}},\ \bibinfo {pages} {1}
  (\bibinfo {year} {2015})},\ \Eprint {http://arxiv.org/abs/1407.0059}
  {arXiv:1407.0059 [astro-ph.CO]} \BibitemShut {NoStop}%
\bibitem [{\citenamefont {Llinares}\ and\ \citenamefont
  {Pogosian}(2014)}]{Llinares:2014zxa}%
  \BibitemOpen
  \bibfield  {author} {\bibinfo {author} {\bibfnamefont {C.}~\bibnamefont
  {Llinares}}\ and\ \bibinfo {author} {\bibfnamefont {L.}~\bibnamefont
  {Pogosian}},\ }\href {\doibase 10.1103/PhysRevD.90.124041} {\bibfield
  {journal} {\bibinfo  {journal} {Phys. Rev.}\ }\textbf {\bibinfo {volume}
  {D90}},\ \bibinfo {pages} {124041} (\bibinfo {year} {2014})},\ \Eprint
  {http://arxiv.org/abs/1410.2857} {arXiv:1410.2857 [astro-ph.CO]} \BibitemShut
  {NoStop}%
\bibitem [{\citenamefont {Gronke}\ \emph {et~al.}(2015)\citenamefont {Gronke},
  \citenamefont {Mota},\ and\ \citenamefont {Winther}}]{Gronke:2015ama}%
  \BibitemOpen
  \bibfield  {author} {\bibinfo {author} {\bibfnamefont {M.}~\bibnamefont
  {Gronke}}, \bibinfo {author} {\bibfnamefont {D.~F.}\ \bibnamefont {Mota}}, \
  and\ \bibinfo {author} {\bibfnamefont {H.~A.}\ \bibnamefont {Winther}},\
  }\href {\doibase 10.1051/0004-6361/201526611} {\bibfield  {journal} {\bibinfo
   {journal} {Astron. Astrophys.}\ }\textbf {\bibinfo {volume} {583}},\
  \bibinfo {pages} {A123} (\bibinfo {year} {2015})},\ \Eprint
  {http://arxiv.org/abs/1505.07129} {arXiv:1505.07129 [astro-ph.CO]}
  \BibitemShut {NoStop}%
\bibitem [{\citenamefont {Carlesi}\ \emph {et~al.}(2017)\citenamefont
  {Carlesi}, \citenamefont {Mota},\ and\ \citenamefont
  {Winther}}]{Carlesi:2016yas}%
  \BibitemOpen
  \bibfield  {author} {\bibinfo {author} {\bibfnamefont {E.}~\bibnamefont
  {Carlesi}}, \bibinfo {author} {\bibfnamefont {D.~F.}\ \bibnamefont {Mota}}, \
  and\ \bibinfo {author} {\bibfnamefont {H.~A.}\ \bibnamefont {Winther}},\
  }\href {\doibase 10.1093/mnras/stx056} {\bibfield  {journal} {\bibinfo
  {journal} {Mon. Not. Roy. Astron. Soc.}\ }\textbf {\bibinfo {volume} {466}},\
  \bibinfo {pages} {4813} (\bibinfo {year} {2017})},\ \Eprint
  {http://arxiv.org/abs/1612.07053} {arXiv:1612.07053 [astro-ph.CO]}
  \BibitemShut {NoStop}%
\bibitem [{\citenamefont {Gronke}\ \emph {et~al.}(2016)\citenamefont {Gronke},
  \citenamefont {Hammami}, \citenamefont {Mota},\ and\ \citenamefont
  {Winther}}]{Gronke:2016lfd}%
  \BibitemOpen
  \bibfield  {author} {\bibinfo {author} {\bibfnamefont {M.}~\bibnamefont
  {Gronke}}, \bibinfo {author} {\bibfnamefont {A.}~\bibnamefont {Hammami}},
  \bibinfo {author} {\bibfnamefont {D.~F.}\ \bibnamefont {Mota}}, \ and\
  \bibinfo {author} {\bibfnamefont {H.~A.}\ \bibnamefont {Winther}},\ }\href
  {\doibase 10.1051/0004-6361/201628644} {\bibfield  {journal} {\bibinfo
  {journal} {Astron. Astrophys.}\ }\textbf {\bibinfo {volume} {595}},\ \bibinfo
  {pages} {A78} (\bibinfo {year} {2016})},\ \Eprint
  {http://arxiv.org/abs/1609.02937} {arXiv:1609.02937 [astro-ph.CO]}
  \BibitemShut {NoStop}%
\bibitem [{\citenamefont {Voivodic}\ \emph {et~al.}(2017)\citenamefont
  {Voivodic}, \citenamefont {Lima}, \citenamefont {Llinares},\ and\
  \citenamefont {Mota}}]{Voivodic:2016kog}%
  \BibitemOpen
  \bibfield  {author} {\bibinfo {author} {\bibfnamefont {R.}~\bibnamefont
  {Voivodic}}, \bibinfo {author} {\bibfnamefont {M.}~\bibnamefont {Lima}},
  \bibinfo {author} {\bibfnamefont {C.}~\bibnamefont {Llinares}}, \ and\
  \bibinfo {author} {\bibfnamefont {D.~F.}\ \bibnamefont {Mota}},\ }\href
  {\doibase 10.1103/PhysRevD.95.024018} {\bibfield  {journal} {\bibinfo
  {journal} {Phys. Rev.}\ }\textbf {\bibinfo {volume} {D95}},\ \bibinfo {pages}
  {024018} (\bibinfo {year} {2017})},\ \Eprint
  {http://arxiv.org/abs/1609.02544} {arXiv:1609.02544 [astro-ph.CO]}
  \BibitemShut {NoStop}%
\bibitem [{\citenamefont {Desmond}\ \emph {et~al.}(2018)\citenamefont
  {Desmond}, \citenamefont {Ferreira}, \citenamefont {Lavaux},\ and\
  \citenamefont {Jasche}}]{Desmond:2018euk}%
  \BibitemOpen
  \bibfield  {author} {\bibinfo {author} {\bibfnamefont {H.}~\bibnamefont
  {Desmond}}, \bibinfo {author} {\bibfnamefont {P.~G.}\ \bibnamefont
  {Ferreira}}, \bibinfo {author} {\bibfnamefont {G.}~\bibnamefont {Lavaux}}, \
  and\ \bibinfo {author} {\bibfnamefont {J.}~\bibnamefont {Jasche}},\
  }\href@noop {} {\  (\bibinfo {year} {2018})},\ \Eprint
  {http://arxiv.org/abs/1802.07206} {arXiv:1802.07206 [astro-ph.CO]}
  \BibitemShut {NoStop}%
\bibitem [{\citenamefont {Ellewsen}\ \emph {et~al.}(2018)\citenamefont
  {Ellewsen}, \citenamefont {Falck},\ and\ \citenamefont
  {Mota}}]{Ellewsen:2018tww}%
  \BibitemOpen
  \bibfield  {author} {\bibinfo {author} {\bibfnamefont {T.~A.~S.}\
  \bibnamefont {Ellewsen}}, \bibinfo {author} {\bibfnamefont {B.}~\bibnamefont
  {Falck}}, \ and\ \bibinfo {author} {\bibfnamefont {D.~F.}\ \bibnamefont
  {Mota}},\ }\href@noop {} {\  (\bibinfo {year} {2018})},\ \Eprint
  {http://arxiv.org/abs/1803.04197} {arXiv:1803.04197 [astro-ph.CO]}
  \BibitemShut {NoStop}%
\bibitem [{\citenamefont {Mota}(2018)}]{Mota:2018kid}%
  \BibitemOpen
  \bibfield  {author} {\bibinfo {author} {\bibfnamefont {D.~F.}\ \bibnamefont
  {Mota}},\ }\href {\doibase 10.1142/S0218271818300033} {\bibfield  {journal}
  {\bibinfo  {journal} {Int. J. Mod. Phys.}\ }\textbf {\bibinfo {volume}
  {D27}},\ \bibinfo {pages} {1830003} (\bibinfo {year} {2018})}\BibitemShut
  {NoStop}%
\bibitem [{\citenamefont {Burrage}\ \emph {et~al.}(2017)\citenamefont
  {Burrage}, \citenamefont {Copeland},\ and\ \citenamefont
  {Millington}}]{Burrage:2016yjm}%
  \BibitemOpen
  \bibfield  {author} {\bibinfo {author} {\bibfnamefont {C.}~\bibnamefont
  {Burrage}}, \bibinfo {author} {\bibfnamefont {E.~J.}\ \bibnamefont
  {Copeland}}, \ and\ \bibinfo {author} {\bibfnamefont {P.}~\bibnamefont
  {Millington}},\ }\href {\doibase 10.1103/PhysRevD.95.064050,
  10.1103/PhysRevD.95.129902} {\bibfield  {journal} {\bibinfo  {journal} {Phys.
  Rev.}\ }\textbf {\bibinfo {volume} {D95}},\ \bibinfo {pages} {064050}
  (\bibinfo {year} {2017})},\ \bibinfo {note} {[Erratum: Phys.
  Rev.D95,no.12,129902(2017)]},\ \Eprint {http://arxiv.org/abs/1610.07529}
  {arXiv:1610.07529 [astro-ph.CO]} \BibitemShut {NoStop}%
\bibitem [{\citenamefont {McGaugh}\ \emph {et~al.}(2016)\citenamefont
  {McGaugh}, \citenamefont {Lelli},\ and\ \citenamefont
  {Schombert}}]{McGaugh:2016leg}%
  \BibitemOpen
  \bibfield  {author} {\bibinfo {author} {\bibfnamefont {S.}~\bibnamefont
  {McGaugh}}, \bibinfo {author} {\bibfnamefont {F.}~\bibnamefont {Lelli}}, \
  and\ \bibinfo {author} {\bibfnamefont {J.}~\bibnamefont {Schombert}},\ }\href
  {\doibase 10.1103/PhysRevLett.117.201101} {\bibfield  {journal} {\bibinfo
  {journal} {Phys. Rev. Lett.}\ }\textbf {\bibinfo {volume} {117}},\ \bibinfo
  {pages} {201101} (\bibinfo {year} {2016})},\ \Eprint
  {http://arxiv.org/abs/1609.05917} {arXiv:1609.05917 [astro-ph.GA]}
  \BibitemShut {NoStop}%
\bibitem [{\citenamefont {Llinares}(2018)}]{Llinares:2018dtu}%
  \BibitemOpen
  \bibfield  {author} {\bibinfo {author} {\bibfnamefont {C.}~\bibnamefont
  {Llinares}},\ }\href {\doibase 10.1093/mnrasl/sly021} {\  (\bibinfo {year}
  {2018}),\ 10.1093/mnrasl/sly021},\ \Eprint {http://arxiv.org/abs/1802.02001}
  {arXiv:1802.02001 [astro-ph.CO]} \BibitemShut {NoStop}%
\bibitem [{\citenamefont {Hagala}\ \emph {et~al.}(2017)\citenamefont {Hagala},
  \citenamefont {Llinares},\ and\ \citenamefont {Mota}}]{Hagala:2016fks}%
  \BibitemOpen
  \bibfield  {author} {\bibinfo {author} {\bibfnamefont {R.}~\bibnamefont
  {Hagala}}, \bibinfo {author} {\bibfnamefont {C.}~\bibnamefont {Llinares}}, \
  and\ \bibinfo {author} {\bibfnamefont {D.~F.}\ \bibnamefont {Mota}},\ }\href
  {\doibase 10.1103/PhysRevLett.118.101301} {\bibfield  {journal} {\bibinfo
  {journal} {Phys. Rev. Lett.}\ }\textbf {\bibinfo {volume} {118}},\ \bibinfo
  {pages} {101301} (\bibinfo {year} {2017})},\ \Eprint
  {http://arxiv.org/abs/1607.02600} {arXiv:1607.02600 [astro-ph.CO]}
  \BibitemShut {NoStop}%
\bibitem [{\citenamefont {Ip}\ and\ \citenamefont
  {Schmidt}(2018)}]{Ip:2018nhl}%
  \BibitemOpen
  \bibfield  {author} {\bibinfo {author} {\bibfnamefont {H.~Y.~S.}\
  \bibnamefont {Ip}}\ and\ \bibinfo {author} {\bibfnamefont {F.}~\bibnamefont
  {Schmidt}},\ }\href@noop {} {\  (\bibinfo {year} {2018})},\ \Eprint
  {http://arxiv.org/abs/1801.07453} {arXiv:1801.07453 [gr-qc]} \BibitemShut
  {NoStop}%
\bibitem [{\citenamefont {Milgrom}(1983)}]{Milgrom:1983pn}%
  \BibitemOpen
  \bibfield  {author} {\bibinfo {author} {\bibfnamefont {M.}~\bibnamefont
  {Milgrom}},\ }\href {\doibase 10.1086/161131} {\bibfield  {journal} {\bibinfo
   {journal} {Astrophys. J.}\ }\textbf {\bibinfo {volume} {270}},\ \bibinfo
  {pages} {371} (\bibinfo {year} {1983})}\BibitemShut {NoStop}%
\bibitem [{\citenamefont {Bugg}(2015)}]{Bugg:2014bka}%
  \BibitemOpen
  \bibfield  {author} {\bibinfo {author} {\bibfnamefont {D.~V.}\ \bibnamefont
  {Bugg}},\ }\href {\doibase 10.1139/cjp-2014-0057} {\bibfield  {journal}
  {\bibinfo  {journal} {Can. J. Phys.}\ }\textbf {\bibinfo {volume} {93}},\
  \bibinfo {pages} {119} (\bibinfo {year} {2015})},\ \Eprint
  {http://arxiv.org/abs/1405.1695} {arXiv:1405.1695 [physics.gen-ph]}
  \BibitemShut {NoStop}%
\bibitem [{\citenamefont {Skordis}(2009)}]{Skordis:2009bf}%
  \BibitemOpen
  \bibfield  {author} {\bibinfo {author} {\bibfnamefont {C.}~\bibnamefont
  {Skordis}},\ }\href {\doibase 10.1088/0264-9381/26/14/143001} {\bibfield
  {journal} {\bibinfo  {journal} {Class. Quant. Grav.}\ }\textbf {\bibinfo
  {volume} {26}},\ \bibinfo {pages} {143001} (\bibinfo {year} {2009})},\
  \Eprint {http://arxiv.org/abs/0903.3602} {arXiv:0903.3602 [astro-ph.CO]}
  \BibitemShut {NoStop}%
\bibitem [{\citenamefont {Moffat}(2006)}]{Moffat:2005si}%
  \BibitemOpen
  \bibfield  {author} {\bibinfo {author} {\bibfnamefont {J.~W.}\ \bibnamefont
  {Moffat}},\ }\href {\doibase 10.1088/1475-7516/2006/03/004} {\bibfield
  {journal} {\bibinfo  {journal} {JCAP}\ }\textbf {\bibinfo {volume} {0603}},\
  \bibinfo {pages} {004} (\bibinfo {year} {2006})},\ \Eprint
  {http://arxiv.org/abs/gr-qc/0506021} {arXiv:gr-qc/0506021 [gr-qc]}
  \BibitemShut {NoStop}%
\bibitem [{\citenamefont {Moffat}\ and\ \citenamefont
  {Rahvar}(2013)}]{Moffat:2013sja}%
  \BibitemOpen
  \bibfield  {author} {\bibinfo {author} {\bibfnamefont {J.~W.}\ \bibnamefont
  {Moffat}}\ and\ \bibinfo {author} {\bibfnamefont {S.}~\bibnamefont
  {Rahvar}},\ }\href {\doibase 10.1093/mnras/stt1670} {\bibfield  {journal}
  {\bibinfo  {journal} {Mon. Not. Roy. Astron. Soc.}\ }\textbf {\bibinfo
  {volume} {436}},\ \bibinfo {pages} {1439} (\bibinfo {year} {2013})},\ \Eprint
  {http://arxiv.org/abs/1306.6383} {arXiv:1306.6383 [astro-ph.GA]} \BibitemShut
  {NoStop}%
\bibitem [{\citenamefont {Khoury}(2015)}]{Khoury:2014tka}%
  \BibitemOpen
  \bibfield  {author} {\bibinfo {author} {\bibfnamefont {J.}~\bibnamefont
  {Khoury}},\ }\href {\doibase 10.1103/PhysRevD.91.024022} {\bibfield
  {journal} {\bibinfo  {journal} {Phys. Rev.}\ }\textbf {\bibinfo {volume}
  {D91}},\ \bibinfo {pages} {024022} (\bibinfo {year} {2015})},\ \Eprint
  {http://arxiv.org/abs/1409.0012} {arXiv:1409.0012 [hep-th]} \BibitemShut
  {NoStop}%
\bibitem [{\citenamefont {Burrage}\ \emph
  {et~al.}(2016{\natexlab{a}})\citenamefont {Burrage}, \citenamefont
  {Copeland},\ and\ \citenamefont {Millington}}]{Burrage:2016xzz}%
  \BibitemOpen
  \bibfield  {author} {\bibinfo {author} {\bibfnamefont {C.}~\bibnamefont
  {Burrage}}, \bibinfo {author} {\bibfnamefont {E.~J.}\ \bibnamefont
  {Copeland}}, \ and\ \bibinfo {author} {\bibfnamefont {P.}~\bibnamefont
  {Millington}},\ }\href {\doibase 10.1103/PhysRevLett.117.211102} {\bibfield
  {journal} {\bibinfo  {journal} {Phys. Rev. Lett.}\ }\textbf {\bibinfo
  {volume} {117}},\ \bibinfo {pages} {211102} (\bibinfo {year}
  {2016}{\natexlab{a}})},\ \Eprint {http://arxiv.org/abs/1604.06051}
  {arXiv:1604.06051 [gr-qc]} \BibitemShut {NoStop}%
\bibitem [{\citenamefont {Burrage}\ \emph {et~al.}(2018)\citenamefont
  {Burrage}, \citenamefont {Copeland}, \citenamefont {Millington},\ and\
  \citenamefont {Spannowsky}}]{Burrage:2018dvt}%
  \BibitemOpen
  \bibfield  {author} {\bibinfo {author} {\bibfnamefont {C.}~\bibnamefont
  {Burrage}}, \bibinfo {author} {\bibfnamefont {E.~J.}\ \bibnamefont
  {Copeland}}, \bibinfo {author} {\bibfnamefont {P.}~\bibnamefont
  {Millington}}, \ and\ \bibinfo {author} {\bibfnamefont {M.}~\bibnamefont
  {Spannowsky}},\ }\href@noop {} {\  (\bibinfo {year} {2018})},\ \Eprint
  {http://arxiv.org/abs/1804.07180} {arXiv:1804.07180 [hep-th]} \BibitemShut
  {NoStop}%
\bibitem [{\citenamefont {Herranen}\ \emph {et~al.}(2014)\citenamefont
  {Herranen}, \citenamefont {Markkanen}, \citenamefont {Nurmi},\ and\
  \citenamefont {Rajantie}}]{Herranen:2014cua}%
  \BibitemOpen
  \bibfield  {author} {\bibinfo {author} {\bibfnamefont {M.}~\bibnamefont
  {Herranen}}, \bibinfo {author} {\bibfnamefont {T.}~\bibnamefont {Markkanen}},
  \bibinfo {author} {\bibfnamefont {S.}~\bibnamefont {Nurmi}}, \ and\ \bibinfo
  {author} {\bibfnamefont {A.}~\bibnamefont {Rajantie}},\ }\href {\doibase
  10.1103/PhysRevLett.113.211102} {\bibfield  {journal} {\bibinfo  {journal}
  {Phys. Rev. Lett.}\ }\textbf {\bibinfo {volume} {113}},\ \bibinfo {pages}
  {211102} (\bibinfo {year} {2014})},\ \Eprint {http://arxiv.org/abs/1407.3141}
  {arXiv:1407.3141 [hep-ph]} \BibitemShut {NoStop}%
\bibitem [{\citenamefont {Bienayme}\ \emph {et~al.}(2009)\citenamefont
  {Bienayme}, \citenamefont {Famaey}, \citenamefont {Wu}, \citenamefont
  {Zhao},\ and\ \citenamefont {Aubert}}]{Bienayme:2009wb}%
  \BibitemOpen
  \bibfield  {author} {\bibinfo {author} {\bibfnamefont {O.}~\bibnamefont
  {Bienayme}}, \bibinfo {author} {\bibfnamefont {B.}~\bibnamefont {Famaey}},
  \bibinfo {author} {\bibfnamefont {X.}~\bibnamefont {Wu}}, \bibinfo {author}
  {\bibfnamefont {H.~S.}\ \bibnamefont {Zhao}}, \ and\ \bibinfo {author}
  {\bibfnamefont {D.}~\bibnamefont {Aubert}},\ }\href {\doibase
  10.1051/0004-6361/200809978} {\bibfield  {journal} {\bibinfo  {journal}
  {Astron. Astrophys.}\ }\textbf {\bibinfo {volume} {500}},\ \bibinfo {pages}
  {801} (\bibinfo {year} {2009})},\ \Eprint {http://arxiv.org/abs/0904.3893}
  {arXiv:0904.3893 [astro-ph.GA]} \BibitemShut {NoStop}%
\bibitem [{\citenamefont {Margalit}\ and\ \citenamefont
  {Shaviv}(2016)}]{Margalit:2015zla}%
  \BibitemOpen
  \bibfield  {author} {\bibinfo {author} {\bibfnamefont {B.}~\bibnamefont
  {Margalit}}\ and\ \bibinfo {author} {\bibfnamefont {N.~J.}\ \bibnamefont
  {Shaviv}},\ }\href {\doibase 10.1093/mnras/stv2721} {\bibfield  {journal}
  {\bibinfo  {journal} {Mon. Not. Roy. Astron. Soc.}\ }\textbf {\bibinfo
  {volume} {456}},\ \bibinfo {pages} {1163} (\bibinfo {year} {2016})},\ \Eprint
  {http://arxiv.org/abs/1505.04790} {arXiv:1505.04790 [astro-ph.GA]}
  \BibitemShut {NoStop}%
\bibitem [{\citenamefont {Bertotti}\ \emph {et~al.}(2003)\citenamefont
  {Bertotti}, \citenamefont {Iess},\ and\ \citenamefont
  {Tortora}}]{Bertotti:2003rm}%
  \BibitemOpen
  \bibfield  {author} {\bibinfo {author} {\bibfnamefont {B.}~\bibnamefont
  {Bertotti}}, \bibinfo {author} {\bibfnamefont {L.}~\bibnamefont {Iess}}, \
  and\ \bibinfo {author} {\bibfnamefont {P.}~\bibnamefont {Tortora}},\ }\href
  {\doibase 10.1038/nature01997} {\bibfield  {journal} {\bibinfo  {journal}
  {Nature}\ }\textbf {\bibinfo {volume} {425}},\ \bibinfo {pages} {374}
  (\bibinfo {year} {2003})}\BibitemShut {NoStop}%
\bibitem [{\citenamefont {Khoury}\ and\ \citenamefont
  {Weltman}(2004{\natexlab{a}})}]{Khoury:2003aq}%
  \BibitemOpen
  \bibfield  {author} {\bibinfo {author} {\bibfnamefont {J.}~\bibnamefont
  {Khoury}}\ and\ \bibinfo {author} {\bibfnamefont {A.}~\bibnamefont
  {Weltman}},\ }\href {\doibase 10.1103/PhysRevLett.93.171104} {\bibfield
  {journal} {\bibinfo  {journal} {Phys. Rev. Lett.}\ }\textbf {\bibinfo
  {volume} {93}},\ \bibinfo {pages} {171104} (\bibinfo {year}
  {2004}{\natexlab{a}})},\ \Eprint {http://arxiv.org/abs/astro-ph/0309300}
  {arXiv:astro-ph/0309300 [astro-ph]} \BibitemShut {NoStop}%
\bibitem [{\citenamefont {Khoury}\ and\ \citenamefont
  {Weltman}(2004{\natexlab{b}})}]{Khoury:2003rn}%
  \BibitemOpen
  \bibfield  {author} {\bibinfo {author} {\bibfnamefont {J.}~\bibnamefont
  {Khoury}}\ and\ \bibinfo {author} {\bibfnamefont {A.}~\bibnamefont
  {Weltman}},\ }\href {\doibase 10.1103/PhysRevD.69.044026} {\bibfield
  {journal} {\bibinfo  {journal} {Phys. Rev.}\ }\textbf {\bibinfo {volume}
  {D69}},\ \bibinfo {pages} {044026} (\bibinfo {year} {2004}{\natexlab{b}})},\
  \Eprint {http://arxiv.org/abs/astro-ph/0309411} {arXiv:astro-ph/0309411
  [astro-ph]} \BibitemShut {NoStop}%
\bibitem [{\citenamefont {Vainshtein}(1972)}]{Vainshtein:1972sx}%
  \BibitemOpen
  \bibfield  {author} {\bibinfo {author} {\bibfnamefont {A.~I.}\ \bibnamefont
  {Vainshtein}},\ }\href {\doibase 10.1016/0370-2693(72)90147-5} {\bibfield
  {journal} {\bibinfo  {journal} {Phys. Lett.}\ }\textbf {\bibinfo {volume}
  {39B}},\ \bibinfo {pages} {393} (\bibinfo {year} {1972})}\BibitemShut
  {NoStop}%
\bibitem [{\citenamefont {Burrage}\ \emph
  {et~al.}(2016{\natexlab{b}})\citenamefont {Burrage}, \citenamefont
  {Kuribayashi-Coleman}, \citenamefont {Stevenson},\ and\ \citenamefont
  {Thrussell}}]{Burrage:2016rkv}%
  \BibitemOpen
  \bibfield  {author} {\bibinfo {author} {\bibfnamefont {C.}~\bibnamefont
  {Burrage}}, \bibinfo {author} {\bibfnamefont {A.}~\bibnamefont
  {Kuribayashi-Coleman}}, \bibinfo {author} {\bibfnamefont {J.}~\bibnamefont
  {Stevenson}}, \ and\ \bibinfo {author} {\bibfnamefont {B.}~\bibnamefont
  {Thrussell}},\ }\href {\doibase 10.1088/1475-7516/2016/12/041} {\bibfield
  {journal} {\bibinfo  {journal} {JCAP}\ }\textbf {\bibinfo {volume} {1612}},\
  \bibinfo {pages} {041} (\bibinfo {year} {2016}{\natexlab{b}})},\ \Eprint
  {http://arxiv.org/abs/1609.09275} {arXiv:1609.09275 [astro-ph.CO]}
  \BibitemShut {NoStop}%
\bibitem [{\citenamefont {Brax}\ and\ \citenamefont
  {Davis}(2016)}]{Brax:2016wjk}%
  \BibitemOpen
  \bibfield  {author} {\bibinfo {author} {\bibfnamefont {P.}~\bibnamefont
  {Brax}}\ and\ \bibinfo {author} {\bibfnamefont {A.-C.}\ \bibnamefont
  {Davis}},\ }\href {\doibase 10.1103/PhysRevD.94.104069} {\bibfield  {journal}
  {\bibinfo  {journal} {Phys. Rev.}\ }\textbf {\bibinfo {volume} {D94}},\
  \bibinfo {pages} {104069} (\bibinfo {year} {2016})},\ \Eprint
  {http://arxiv.org/abs/1609.09242} {arXiv:1609.09242 [astro-ph.CO]}
  \BibitemShut {NoStop}%
\bibitem [{\citenamefont {Jaffe}\ \emph {et~al.}(2017)\citenamefont {Jaffe},
  \citenamefont {Haslinger}, \citenamefont {Xu}, \citenamefont {Hamilton},
  \citenamefont {Upadhye}, \citenamefont {Elder}, \citenamefont {Khoury},\ and\
  \citenamefont {Müller}}]{Jaffe:2016fsh}%
  \BibitemOpen
  \bibfield  {author} {\bibinfo {author} {\bibfnamefont {M.}~\bibnamefont
  {Jaffe}}, \bibinfo {author} {\bibfnamefont {P.}~\bibnamefont {Haslinger}},
  \bibinfo {author} {\bibfnamefont {V.}~\bibnamefont {Xu}}, \bibinfo {author}
  {\bibfnamefont {P.}~\bibnamefont {Hamilton}}, \bibinfo {author}
  {\bibfnamefont {A.}~\bibnamefont {Upadhye}}, \bibinfo {author} {\bibfnamefont
  {B.}~\bibnamefont {Elder}}, \bibinfo {author} {\bibfnamefont
  {J.}~\bibnamefont {Khoury}}, \ and\ \bibinfo {author} {\bibfnamefont
  {H.}~\bibnamefont {Müller}},\ }\href {\doibase 10.1038/nphys4189} {\bibfield
   {journal} {\bibinfo  {journal} {Nature Phys.}\ }\textbf {\bibinfo {volume}
  {13}},\ \bibinfo {pages} {938} (\bibinfo {year} {2017})},\ \Eprint
  {http://arxiv.org/abs/1612.05171} {arXiv:1612.05171 [physics.atom-ph]}
  \BibitemShut {NoStop}%
\bibitem [{\citenamefont {Will}(2006)}]{Will:2005va}%
  \BibitemOpen
  \bibfield  {author} {\bibinfo {author} {\bibfnamefont {C.~M.}\ \bibnamefont
  {Will}},\ }\href {\doibase 10.12942/lrr-2006-3} {\bibfield  {journal}
  {\bibinfo  {journal} {Living Rev. Rel.}\ }\textbf {\bibinfo {volume} {9}},\
  \bibinfo {pages} {3} (\bibinfo {year} {2006})},\ \Eprint
  {http://arxiv.org/abs/gr-qc/0510072} {arXiv:gr-qc/0510072 [gr-qc]}
  \BibitemShut {NoStop}%
\bibitem [{\citenamefont {Sakstein}(2018)}]{Sakstein:2017pqi}%
  \BibitemOpen
  \bibfield  {author} {\bibinfo {author} {\bibfnamefont {J.}~\bibnamefont
  {Sakstein}},\ }\href {\doibase 10.1103/PhysRevD.97.064028} {\bibfield
  {journal} {\bibinfo  {journal} {Phys. Rev.}\ }\textbf {\bibinfo {volume}
  {D97}},\ \bibinfo {pages} {064028} (\bibinfo {year} {2018})},\ \Eprint
  {http://arxiv.org/abs/1710.03156} {arXiv:1710.03156 [astro-ph.CO]}
  \BibitemShut {NoStop}%
\bibitem [{\citenamefont {Read}(2014)}]{Read:2014qva}%
  \BibitemOpen
  \bibfield  {author} {\bibinfo {author} {\bibfnamefont {J.~I.}\ \bibnamefont
  {Read}},\ }\href {\doibase 10.1088/0954-3899/41/6/063101} {\bibfield
  {journal} {\bibinfo  {journal} {J. Phys.}\ }\textbf {\bibinfo {volume}
  {G41}},\ \bibinfo {pages} {063101} (\bibinfo {year} {2014})},\ \Eprint
  {http://arxiv.org/abs/1404.1938} {arXiv:1404.1938 [astro-ph.GA]} \BibitemShut
  {NoStop}%
\bibitem [{\citenamefont {Silverwood}\ \emph {et~al.}(2016)\citenamefont
  {Silverwood}, \citenamefont {Sivertsson}, \citenamefont {Steger},
  \citenamefont {Read},\ and\ \citenamefont {Bertone}}]{Silverwood:2015hxa}%
  \BibitemOpen
  \bibfield  {author} {\bibinfo {author} {\bibfnamefont {H.}~\bibnamefont
  {Silverwood}}, \bibinfo {author} {\bibfnamefont {S.}~\bibnamefont
  {Sivertsson}}, \bibinfo {author} {\bibfnamefont {P.}~\bibnamefont {Steger}},
  \bibinfo {author} {\bibfnamefont {J.~I.}\ \bibnamefont {Read}}, \ and\
  \bibinfo {author} {\bibfnamefont {G.}~\bibnamefont {Bertone}},\ }\href
  {\doibase 10.1093/mnras/stw917} {\bibfield  {journal} {\bibinfo  {journal}
  {Mon. Not. Roy. Astron. Soc.}\ }\textbf {\bibinfo {volume} {459}},\ \bibinfo
  {pages} {4191} (\bibinfo {year} {2016})},\ \Eprint
  {http://arxiv.org/abs/1507.08581} {arXiv:1507.08581 [astro-ph.GA]}
  \BibitemShut {NoStop}%
\bibitem [{\citenamefont {Flynn}\ \emph {et~al.}(2006)\citenamefont {Flynn},
  \citenamefont {Holmberg}, \citenamefont {Portinari}, \citenamefont {Fuchs},\
  and\ \citenamefont {Jahreiss}}]{Flynn:2006tm}%
  \BibitemOpen
  \bibfield  {author} {\bibinfo {author} {\bibfnamefont {C.}~\bibnamefont
  {Flynn}}, \bibinfo {author} {\bibfnamefont {J.}~\bibnamefont {Holmberg}},
  \bibinfo {author} {\bibfnamefont {L.}~\bibnamefont {Portinari}}, \bibinfo
  {author} {\bibfnamefont {B.}~\bibnamefont {Fuchs}}, \ and\ \bibinfo {author}
  {\bibfnamefont {H.}~\bibnamefont {Jahreiss}},\ }\href {\doibase
  10.1111/j.1365-2966.2006.10911.x} {\bibfield  {journal} {\bibinfo  {journal}
  {Mon. Not. Roy. Astron. Soc.}\ }\textbf {\bibinfo {volume} {372}},\ \bibinfo
  {pages} {1149} (\bibinfo {year} {2006})},\ \Eprint
  {http://arxiv.org/abs/astro-ph/0608193} {arXiv:astro-ph/0608193 [astro-ph]}
  \BibitemShut {NoStop}%
\end{thebibliography}%

\end{document}